\documentclass[twocolumn,english,aps,prd,reprint,floatfix,notitlepage,footinbib,preprintnumbers,superscriptaddress,longbibliography]{revtex4-1}
\pdfoutput=1
\usepackage{lmodern}

\usepackage[T1]{fontenc}
\usepackage[latin9]{inputenc}
\usepackage{geometry}
\usepackage{subfigure,lmodern, amsmath,amssymb, graphicx, pifont, adjustbox, bm, xcolor}
\usepackage{amsfonts}
\usepackage{enumitem}
\usepackage{comment}
\usepackage{mathtools}
\usepackage{float}
\usepackage{slashed}
\usepackage{ragged2e}
\usepackage{array}
\usepackage{bbm}

\usepackage{nameref}

\usepackage{hhline}

\makeatletter\g@addto@macro\bfseries{\boldmath}\makeatother

\makeatletter\newcommand{\labeltext}[2]{%
  \def\@currentlabel{#1}%
  \label{#2}%
}
\makeatother

\usepackage{stackengine}
\usepackage{esint}
\usepackage[unicode=true,pdfusetitle,
 bookmarks=true,bookmarksnumbered=false,bookmarksopen=false,
 breaklinks=false,pdfborder={0 0 1},backref=false,colorlinks=true]
 {hyperref}
\hypersetup{
 pdfauthor={Francesco Calisto, Clifford Cheung, Grant N. Remmen, Francesco Sciotti, Michele Tarquini},
 citecolor=black,linkcolor=black,urlcolor=black}

\newcommand{\appendixref}[1]{\hyperref[#1]{appendix~\ref{#1}}}
\def\equationautorefname~#1\null{eq.\,(#1)\null}
\usepackage{breakurl}
\usepackage{breakurl}
\usepackage[hang,flushmargin]{footmisc} 
\allowdisplaybreaks
\makeatletter

\usepackage{etoolbox}
\apptocmd{\thebibliography}{\justifying\setlength{\leftskip}{7.4mm}}{}{} 
 
 \usepackage{relsize}
\usepackage{babel}

\makeatletter
\def\simgt{\mathrel{\lower2.5pt\vbox{\lineskip=0pt\baselineskip=0pt
           \hbox{$>$}\hbox{$\sim$}}}}
\def\simlt{\mathrel{\lower2.5pt\vbox{\lineskip=0pt\baselineskip=0pt
           \hbox{$<$}\hbox{$\sim$}}}}
\makeatother

\usepackage{changepage}

\newcommand{\be}{\begin{equation}}
\newcommand{\ee}{\end{equation}}
\newcommand{\bea}{\begin{eqnarray}}
\newcommand{\eea}{\end{eqnarray}}
\newcommand{\Fig}[1]{Fig.~\ref{#1}}

\newcommand{\Eq}[1]{Eq.~(\ref{#1})}

\newcommand{\Sec}[1]{Sec.~\ref{#1}}

\newcommand{\App}[1]{App.~\ref{#1}}

\newcommand{\eq}[2]{\be\begin{aligned}#1 \label{#2}\end{aligned}\ee}

\newcommand{\Proj}{\mathcal{P}}

\newcolumntype{P}[1]{>{\centering\arraybackslash}p{#1}}

\usepackage{fix-cm}

\begin{document}

\makeatletter \renewcommand{\l@subsection}[2]{} \makeatother

\preprint{CALT-TH 2025-040}

\title{Completeness from Gravitational Scattering}

\author{Francesco Calisto}
\affiliation{Walter Burke Institute for Theoretical Physics and
Leinweber Forum for Theoretical Physics, California Institute of Technology, Pasadena, CA 91125, USA}
\author{Clifford Cheung}
\affiliation{Walter Burke Institute for Theoretical Physics and
Leinweber Forum for Theoretical Physics, California Institute of Technology, Pasadena, CA 91125, USA}
\author{Grant N.~Remmen}
\affiliation{\scalebox{1}{Center for Cosmology and Particle Physics, Department of Physics, New York University, New York, NY 10003, USA}}    
\author{Francesco Sciotti}
\affiliation{IFAE and BIST, Universitat Aut\`onoma de Barcelona, 08193 Bellaterra, Barcelona, Spain}
\author{Michele Tarquini}
\affiliation{Walter Burke Institute for Theoretical Physics and
Leinweber Forum for Theoretical Physics, California Institute of Technology, Pasadena, CA 91125, USA}

\begin{abstract}

\noindent 
We prove that symmetry in the presence of gravity implies a version of the completeness hypothesis.  For a broad class of theories, we demonstrate that the existence of finitely many charged particles logically necessitates the existence of infinitely many charged particles populating the entire charge lattice.
Our conclusions follow from the consistency of perturbative gravitational scattering and require the following ingredients: 1)~a weakly coupled ultraviolet completion of gravity, 2)~a nonabelian symmetry $G$, gauged or global, whose Cartan subgroup generates the abelian charge lattice, and 3)~a spectrum containing some finite set of charged representations,  in the simplest cases taken to be a single particle in the fundamental.
Under these conditions,  the abelian charge lattice is completely filled by single-particle states for $G=SO(N)$ with $N\geq 5$ and $G=SU(N)$ with $N\geq  3$, which in turn implies completeness for other symmetry groups such as $Spin(N)$, $Sp(N)$, and $E_8$.
Curiously, a corollary of our results is that the $SU(5)$ and $SO(10)$ grand unified theories have precisely the minimal field content needed to derive completeness using our methodology.
\end{abstract}

\maketitle 

\tableofcontents

\section{Introduction}\vspace{-0.7mm}
What are the constituents of the universe?  Ultimately, this is a question to be decided by experiment.  At the same time, it is worth noting that mathematical consistency alone severely limits what can exist in nature, even in principle. For example, Wigner famously showed that the menu of conceivable physical states is not arbitrary, but rigidly constrained by unitarity and special relativity~\cite{Wigner:1939cj}.   Furthermore, modern developments in quantum field theory and scattering amplitudes have established that the perturbative dynamics of particles are almost entirely fixed by their kinematical properties.  The only self-interacting theories of massless particles of spin one and spin two are gauge theory and gravity, while higher-spin massless particles are inconsistent~\cite{Weinberg:1980kq,Benincasa:2007xk,McGady:2013sga,Arkani-Hamed:2017jhn,Elvang:2015rqa,Cheung:2017pzi}.

These insights have demonstrated that certain states are mathematically forbidden. On the other hand, the converse possibility---that certain states might actually be mathematically {\it required}---is equally if not more intriguing. 
The maximalist incarnation of this idea is the notion of {\it completeness}, which is the property that all charges permitted by Dirac quantization are explicitly realized by physical states in the spectrum.  

 It has been conjectured that completeness is a universal feature of all consistent theories of quantum gravity~\cite{Polchinski:2003bq,Banks:2010zn},  referred to collectively as the landscape.  The complement of this space is the swampland, which describes the set of naively sensible gravitational effective field theories that can never actually be realized by any ultraviolet completion \cite{Vafa:2005ui}.
  A well-known motivation for the completeness hypothesis is the absence of global symmetries in quantum gravity~\cite{Banks:1988yz}.  In particular, to explicitly break a global higher-form symmetry, one posits the existence of particles of all allowed charges~\cite{Heidenreich:2021xpr,Rudelius:2020orz}.

In this paper, we adopt an entirely different approach to this question.   Using bottom-up reasoning, we rigorously prove a version of the completeness hypothesis. Our conclusions apply to a variety of theories and are derived purely from the mathematical consistency of scattering amplitudes.  The workhorse of our methodology is the analytic dispersion relation,
\eq{
c_n(t) &= \frac{1}{2 \pi i } \oint_{s=0} \frac{ds}{s^{n+1}} A(s,t) \\
&= \left\{ s\textrm{ channel} \right\} + \left\{ u\textrm{ channel} \right\}+ b_n(t).
}{DR}
At fixed $t$, this expression extracts the Wilson coefficient $c_n(t)$ from the four-point scattering amplitude $A(s,t)$ and relates it to a boundary contribution $b_n(t)$ at infinity plus a sum over discontinuities in the $s$ and $u$ channels.  This equation is the key ingredient of our analysis: if we {\it know} that $c_n(t) \neq 0$ and $b_{n}(t) =0$, then there {\it must} be a state in either the $s$ channel or the $u$ channel, or both.

Physically, the conditions $c_n(t)\neq0$ and $b_n(t)=0$ imply that there is some operator that blows up faster in the effective field theory than in the full amplitude.  This is the usual situation in which the effective field theory dynamics are ``unitarized'' by the ultraviolet completion.  
Crucially, \Eq{DR} is only useful for deducing the existence of states if there is a mandatory coupling $c_n(t)\neq 0$ that is always unitarized, so that $b_n(t)=0$.

\begin{figure}
\includegraphics[width=0.98\columnwidth]{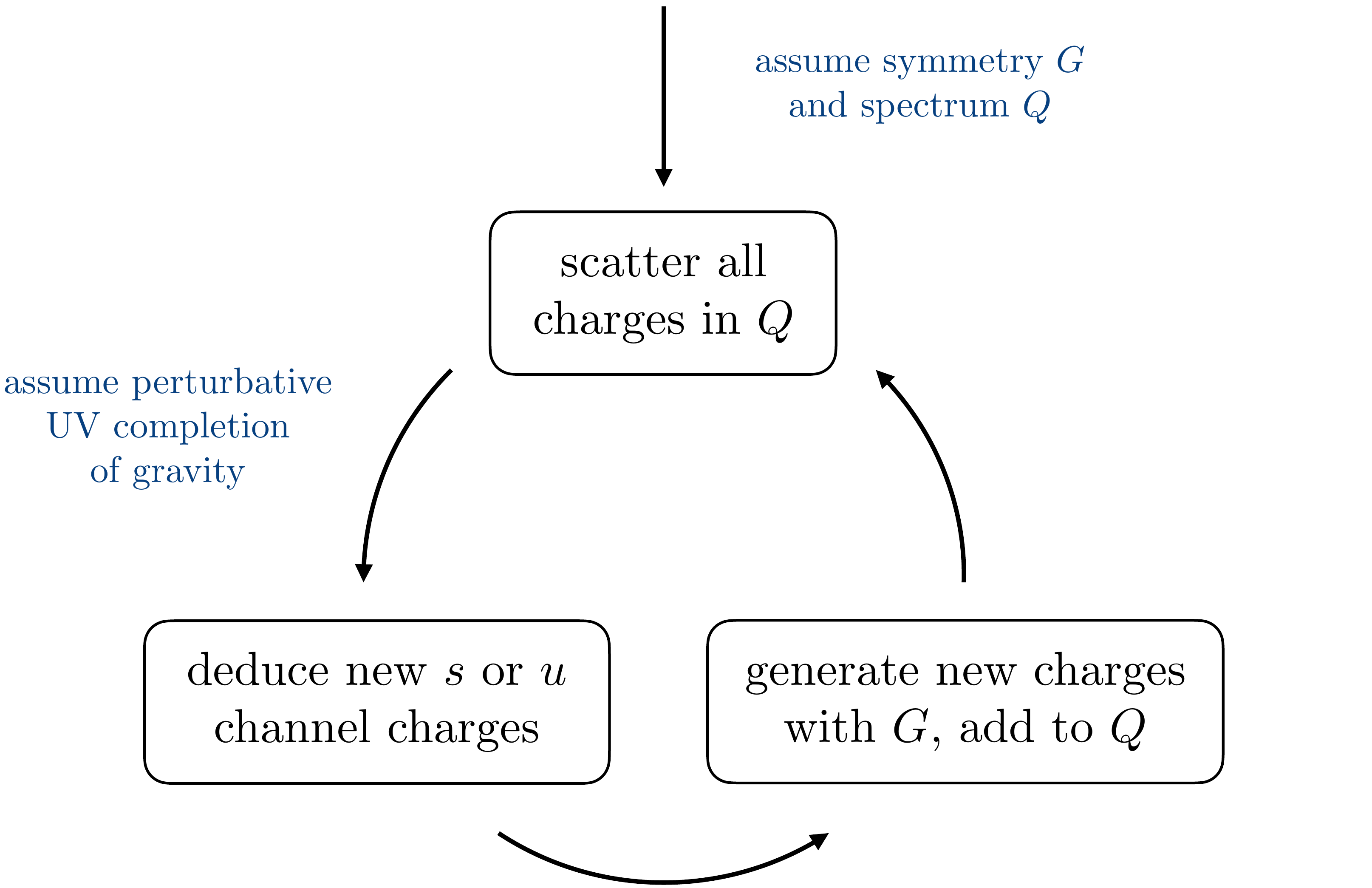}
\caption{Completeness algorithm used to derive charge completeness from gravitational scattering.}
\label{algchart}
\end{figure}

Remarkably, {\it quantum gravity} furnishes the exact conditions that we desire~\cite{Hillman:2024ouy}.  The equivalence principle says that any pair of particles will interact gravitationally. The corresponding graviton exchange contribution appears near the forward limit of the amplitude~\footnote{Here the ellipses denote all terms in the low-energy amplitude that are not the graviton pole. This includes contributions that are regular in $t$ coming from general relativity, as well as higher-dimension operator corrections in the effective field theory.  Notably, all of these terms exhibit different $t$ dependence than the graviton pole, so there are always values of $t$ for which the $s^2$ coefficient of the low-energy amplitude is nonzero, for instance if $t$ is below the mass gap.},
\eq{
A(s,t) = -\frac{8\pi G_N s^2}{t} + \cdots,
}{eq:GREFTamp}
so $c_2(t) =-8\pi G_N /t$ is nonzero for any small but finite $t$.  Notably, it has been argued from the bottom up that $b_2(t)=0$ for $t<0$ under mild assumptions about gravitational scattering~\footnote{In Ref.~\cite{Haring:2022cyf}, Regge bounds were derived for gravitational scattering amplitudes as a consequence of analyticity, unitarity, and the partial wave expansion in various contexts. The most conservative of these bounds, merely assuming dominance of single-graviton exchange at large impact parameter, showed in generality that, for tree-level amplitudes, $A(s,t)<s^{2-\frac{D-7}{2(D-4)}}$ for $t<0$, enabling twice-subtracted dispersion relations in $D>7$.  By including eikonalization, even better scaling is possible. More generally, Refs.~\cite{Arkani-Hamed:2020blm, Caron-Huot:2022ugt} outlined causality arguments that bound $A(s,t)<s^2$.}. 
Furthermore, $b_2(t)=0$ for $t<0$ in all known examples from string theory,
where an infinite tower of higher-spin particles intervenes to unitarize gravitational scattering.  The dispersion relation in \Eq{DR}  then implies that~\cite{Haring:2024wyz}
\eq{
c_2(t) = -\frac{8\pi G_N}{t} = \left\{ s\textrm{ channel} \right\} + \left\{ u\textrm{ channel} \right\},
}{eq:c2su}
so there has to be a state in the $s$ or $u$ channel.

 In the presence of an exact symmetry, \Eq{eq:c2su}  can be used to prove highly nontrivial constraints on the spectrum of corresponding charges~\cite{Hillman:2024ouy}.
In our setup we will assume an exact symmetry group $G$ whose maximal abelian subgroup is $H$.  We take $G$ to be  a finite semisimple Lie algebra, so $H$ is the Cartan subgroup.  Whether $G$ is gauged or not will not matter for our analysis.  

\begin{table}[t]
\centering
\renewcommand{\arraystretch}{1.3}
\begin{tabular}{@{}lll@{}}
\hline\hline
\textbf{Symmetry $G$ \qquad } & \textbf{Spectrum $Q$ \qquad} & \textbf{Section} \\
\hline
$\mathit{SO}(N),\;N\!\geq\!5$               & $\boldsymbol{N}$                                 & \hyperref[sec:SON]{IV\,C--D} \\
$\mathit{SU}(N),\;N\!\geq\!4$               & state of each $N$-ality            & \hyperref[sec:SUN]{V\,B} \\
$\mathit{Spin}(N),\;N\!\geq\!4$             &  $\boldsymbol{N}$ and $\mathbf{2}^{\boldsymbol{\lfloor(N-1)/2\rfloor}}$                        & \hyperref[sec:SpinN]{VI}     \\
$\mathit{Sp}(N),\;N\!\geq\!6$               & $\boldsymbol{2N}$                                & \hyperref[sec:SpN]{VII}    \\
$E_8$                                        & $\mathbf{248}$                               & \hyperref[sec:E8]{VIII}   \\
\hline\hline
\end{tabular}
\caption{Table of the minimal starting spectrum of charges $Q$ required so that our methods successfully derive charge completeness for the group $G$.  Our approach fails for the lower rank groups not listed.  The simplest examples that establish completeness are $SO(5)$ with $\boldsymbol{5}$, and $SU(3)$ with $\mathbf{3}$ and $\mathbf{10}$.   The phenomenologically interesting case of GUTs corresponds to $SU(5)$ with $\mathbf{5}+\mathbf{10}+\mathbf{24}$ and $\mathit{Spin}(10)$ with $\mathbf{10}+\mathbf{16}$.  Completeness is not derived for $SO(4)$ with $\boldsymbol{4}$, but it can be derived if we enlarge to $O(4)$.
}
\label{table}
\end{table}

We then fold \Eq{eq:c2su} into the simple iterative procedure shown schematically in \Fig{algchart}.  In the very first step, we specify some initial set of particles $Q$ that are assumed to be in the spectrum and taken to be in the fundamental of $G$ unless stated otherwise.  We then scatter all pairs of particles within $Q$ and apply \Eq{eq:c2su} in order to deduce the existence of additional charged states.  After adding these new states to $Q$,  we {\it rescatter} all the elements of $Q$ again, taking the newly-deduced particles to be external states of yet another scattering process.  Since these particles also interact gravitationally,  \Eq{eq:c2su} applies once again, and so on ad infinitum.
By iterating this algorithm, we very generically discover that an infinite tower of charged particles is required purely by mathematical consistency.  
 If the spectrum of states eventually grows to encompass the full lattice of possible charges, or weights, under $H$, then we say that the theory exhibits ``charge completeness.''

Remarkably, we are able to prove charge completeness across a broad range of theories, though our precise conclusions depend sensitively on the degree of symmetry in $G$. These criteria are summarized in Table~\ref{table}. For example, for the abelian symmetry group $G = SO(2)\simeq U(1)$, our methods are, unfortunately, insufficient to prove anything. The same is true of $G =  SO(3) \sim SU(2)$ and $SO(4)\sim SO(3)^2$.  However, completeness does follow if we consider larger groups. For $G=SO(N)$ with $N\geq 5$, the additional nonabelian structure accommodates an explicit constructive proof of charge completeness. Meanwhile, for $G=SU(N)$ with $N\geq3$, we derive completeness for various infinite subsets of charges that depend on what precisely we assume for the starting representations.
Our results suggest that spectral completeness is a robust property of any weakly coupled ultraviolet completion of gravity with a sufficiently large nonabelian symmetry and a finite but appropriately chosen initial set of charged particles.   

Our results have implications for the phenomenologically relevant case of grand unified theory (GUT) with $G= SU(5)$ or $Spin(10)$.  
As is well known, a proper embedding of the standard model within $SU(5)$ GUT mandates the existence of the $\mathbf{5}$ and $\mathbf{10}$ for the matter and the $\mathbf{24}$ for the Higgs.  Curiously, this set of representations is precisely sufficient to derive charge completeness, with any smaller set failing to do so.
Meanwhile, $Spin(10)$ GUT requires the $\mathbf{16}$ for the matter and the $\mathbf{10}$ for the Higgs.  In this case, the phenomenologically required matter representations correspond exactly to the minimal set needed to derive charge completeness.  In other words, the phenomenologically required field content of the $SU(5)$ and $Spin(10)$ GUTs is exactly sufficient to guarantee the completeness of electromagnetic charges. 

For clarity, let us tabulate our assumptions very explicitly.  Our arguments rigorously imply a version of the completeness hypothesis, provided there is:
\begin{itemize}[itemsep=0pt]
\renewcommand\labelitemi{}
    \item {\it i}) an exact symmetry $G$,
    \item {\it ii}) a starting set of  charged states $Q$, and
    \item {\it iii}) a tree-level ultraviolet completion of gravity.
\end{itemize}
Obviously, assumptions {\it i}) and {\it ii}) are required just to have a symmetry to speak of and some seed set of states to scatter.  We will elaborate in immense detail later on about the precise choices of $G$ and $Q$ for which completeness is established.

Our final condition {\it iii}) states that the gravitational dynamics are unitarized at tree level.  Mathematically, this implies that graviton scattering is softened at high energies, so for $t<0$ we have  
$b_2(t) = 0$, justifying the application of  \Eq{eq:c2su}.    We emphasize that this last assumption is exceedingly well motivated and conservative. Any ultraviolet completion of gravity that improves the high-energy behavior of general relativity by any amount---which is the very definition of ultraviolet completion---will satisfy this condition.  From the top down, 
this attribute is strongly motivated because it is satisfied by {\it all perturbative string theories}.   This enormous class of models, which happen to be the only extant quantum gravitational theories with explicit and exhaustive predictions for scattering amplitudes, have the universal property that $b_2(t) = 0$.  From the bottom up, the assumption that $b_2(t) = 0$ is also very mild and
supported by general arguments~\cite{Haring:2022cyf,Cheung:2016wjt}.    Notably, the condition of tree-level dynamics, together with tame Regge behavior, is actually sufficient to bootstrap string amplitudes uniquely~\cite{Cheung:2024uhn,Cheung:2024obl, Cheung:2025tbr}.

An important consequence of our assumptions is that the notion of completeness that we establish is both stronger and weaker than the conventional notion of completeness invoked in the swampland literature~\cite{Polchinski:2003bq,Banks:2010zn,Harlow:2018tng,Heidenreich:2021xpr,Rudelius:2020orz}.  In those works, completeness makes no reference to whether the charged states are single-particle or multi-particle.  That is, the presence of ultracharged states is essentially trivialized by the existence of multi-particle states, provided there already exist states of some fundamental charge.  For this reason the weight of the swampland conjectures falls on whether these fundamentally charged states are present in the first place.  
By contrast, our analysis by fiat assumes that some minimally charged states are present, which in a sense weakens our conclusions.  On the other hand, our arguments mandate the existence of ultracharged states that are single particles, which is much stronger than is required by the usual swampland conjectures. 
Perturbative string theory famously exhibits completeness of the spectrum by way of single-particle states, and remarkably, our methodology arrives at the very same conclusion using bootstrap methods.

The remainder of this paper is structured as follows.  We begin in \Sec{sec:completeness_algorithm} by outlining a general iterative procedure for constructively deriving completeness of the charge lattice.  As a warmup, we apply this algorithm to $U(1)$ symmetry  in Sec.~\ref{sec:U1} and explain why completeness cannot be derived in that very simplest case.  Afterwards, in Secs.~\ref{sec:SON} and \ref{sec:SUN} we derive completeness for $SO(N)$ and $SU(N)$ and generalize to $Spin(N)$, $Sp(N)$, and $E_8$ in Secs.~\ref{sec:SpinN}, \ref{sec:SpN}, and \ref{sec:E8}. We then discuss the implications of these results for GUTs in \Sec{sec:GUT} and summarize our conclusions and future directions in \Sec{sec:discussion}.

\section{Completeness Algorithm} \label{sec:completeness_algorithm}

In this section, we outline our constructive procedure for deriving completeness.  
To begin, we initialize the algorithm by specifying the symmetries of the theory, together with some finite set of particles assumed to be in the spectrum.  Since the full spectrum must be invariant under the action of the symmetry, we can actually exploit the symmetry to generate new charges from old ones.  In particular, starting from any given charge we can apply the generators of the symmetry to construct additional families of charges from this state.     We will refer to this action as ``orbiting'' the charge and the set of resulting charges as its ``orbit.''  

We then scatter pairs of particles and apply dispersion relations to deduce the existence of new charges, iterating the algorithm to consider all possible scattering processes for the additional charges that we find.  At each step, we orbit the charges in hand to generate as many new ones as possible.  By repeating this procedure, we incrementally populate the charge lattice.  If the space of charges ultimately spans the full lattice, completeness is established and we claim victory.
The algorithm is summarized in \Fig{algchart}.

\subsection{Initialization}\label{sec:initialization}
We assume throughout that the dynamics are invariant under an exact symmetry described by a group $G$.  The eigenvectors of its Cartan subgroup $H$ span a quantized lattice $\Lambda$ defining the  abelian charges.

Let us define $Q$ to be the set of charges that we know to be in the spectrum at any given point in this procedure.  
At the very start of the algorithm, we initialize $Q$ to be some finite set of charges assumed by fiat to be in the spectrum.  In all cases, $Q$ will include the graviton, which is by definition a singlet under $G$, together with an additional particle that will usually be the fundamental.  In specific cases we may sometimes use a different initial choice of representations.

The spectrum $Q$ at any given step must be invariant under the action of $G$.    Furthermore, we will sometimes encounter outer automorphisms $C$ that are not contained in $G$, but nevertheless leave $Q$ invariant.
Formally, we express these invariances of the spectrum as the statement that $G(Q)=C(Q) = Q$.

It is essential to our argument that we can orbit a particular charge by acting on it with $G$ and $C$ to generate the span of all related charges.     This will be an indispensable trick for generating whole new families of charges from a single seed charge.   At a technical level, the orbit of a charge is implemented by the Weyl symmetries of the root system of $G$, together with the outer automorphisms $C$. 
See \App{groupthreview} for a review of various group theory definitions.

We will discuss the specific mechanics of these orbits in detail when we consider explicit examples.  For the moment, let us simply note that the action of the orbit has a nice pictorial interpretation in terms of points moving about the charge lattice $\Lambda$.  A given charge defines a point in $\Lambda$, and the Weyl symmetry orbits this point through the vertices at the boundary of a polytope in $\Lambda$ corresponding to the highest weight states of a certain representation.  The analogue of lowering operators in $G$ can then  be used to orbit the boundary inwards, filling in swaths of the hull enclosed by the polytope. For reasons we will explain later, the orbit of the original charge carries the same central charge as the seed. 

The mechanics of orbiting a particular charge is of course entirely familiar from the $SU(2)$ description of angular momentum in quantum mechanics. Starting from a state of azimuthal angular momentum $J_z$, we are guaranteed the existence of a whole family of spinning states related by the sequential application of the lowering operator down each rung to $-J_z$.  Alternatively, we can instead apply the Weyl group of $SU(2)$ to jump directly from $J_z$ to $-J_z$.  The latter corresponds to orbiting through the vertices at the boundary of the spin representation, while the former corresponds to orbiting inwards towards the interior.  Note that if we know $J_z$ but we do not know $J$, we are not guaranteed the existence of new states of larger azimuthal angular momentum, since the raising operator can simply yield zero.  This is why we can only orbit inwards and not outwards.

\subsection{Iteration}\label{sec:algorithm}

We are now ready to describe the algorithm for proving completeness.  Our end goal is to grow the spectrum $Q$ at each step.  If at any point we can show that $Q=\Lambda$, then completeness is established.   The steps in this procedure are as follows:
\begin{itemize}[
    labelindent=0pt,       
    leftmargin=0.2cm,        
    rightmargin=0.2cm,       
    labelsep=0.5em,        
    align=left             
]

\item[] 1) Scatter all possible combinations of charged particles.  Every possible process is labeled by a pair of charges $\vec{q}, \vec{q}^{\,\prime} \in Q$.  At the level of the four-point scattering amplitude, the external charges are $\vec q_1 = -\vec{q}_4 = \vec{q}$ and $\vec q_2 = -\vec q_3=\vec{q}^{\,\prime} $.  We require this elastic charge configuration so that the $t$-channel state is neutral.  Only then can the graviton contribute nontrivially to the left-hand side of \Eq{eq:c2su}, which is needed to deduce with certainty the existence of some exchanged state on the right-hand side.

\item[] 2) Apply the dispersion relation in \Eq{eq:c2su} to deduce the existence of a particle either in the $s$ or $u$ channel, 
 \eq{
  \includegraphics[width=0.95\columnwidth]{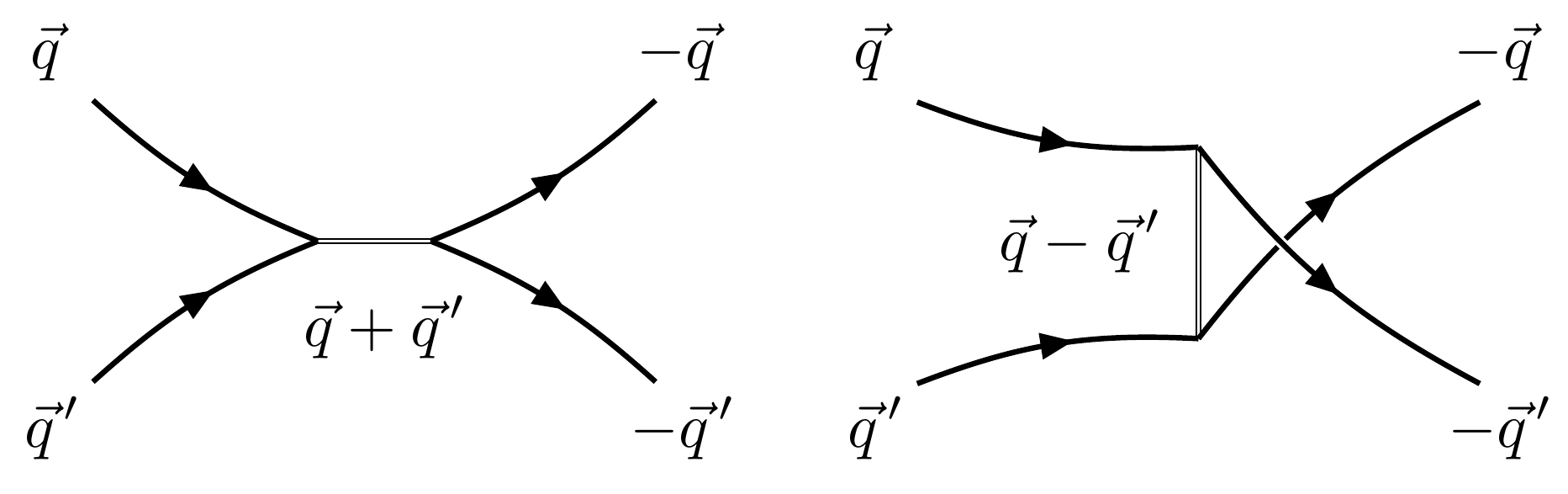} \nonumber
  }{}
\par\vspace{-1em}
whose charges are shown.   We will depict the corresponding logical deduction symbolically as
\eq{
\vec q \otimes  \vec{q}^{\,\prime} \rightarrow ( \vec q +\vec{q}^{\,\prime}) \lor ( \vec q -\vec{q}^{\,\prime}).
}{eq:scattgen}
By $CPT$ invariance, every particle is accompanied by an antiparticle of opposite charge.  Consequently,  the above logic also implies the existence of a state of charge either $ -\vec q -\vec{q}^{\,\prime}$ or $- \vec q + \vec{q}^{\,\prime}$.  For brevity we will not explicitly list these $CPT$ conjugates at each step in the algorithm, since they are automatically present.

\item[] 3)   If both $\vec{q}+ \vec{q}^{\,\prime} \notin Q$ and $\vec{q}- \vec{q}^{\,\prime} \notin Q$, then the dispersion relation in \Eq{eq:c2su} guarantees the existence of a new charge in the spectrum.  We refer to such a scattering process as ``conclusive.''  Conversely, if either $\vec{q}+ \vec{q}^{\,\prime} \in Q$ or $\vec{q}- \vec{q}^{\,\prime} \in Q$, then no new charges are strictly required.  In such a case we deduce nothing, so the scattering process is deemed ``inconclusive.''  

\item[] 4) If any scattering process is conclusive, then we update $Q$ to include the required new charges.  At this step we compute the orbit of the new charges to generate the full space of charges required by the symmetry $G$ and any outer automorphisms $C$.  The precise mechanics of this maneuver will depend on the situation.  In some cases $\vec q + \vec q^{\, \prime}$ and $\vec q - \vec q^{\, \prime}$ will be trivially related to each other by the action of $G$ or $C$, so it does not matter which channel is activated.
In other cases, the charges will be distinct, and we must adapt our strategy accordingly in order to determine which channel's charges must be added. Either way, after we have assembled some set of newly deduced charges, we then append them to $Q$ and return to step 1) to iterate.

\end{itemize}

\noindent If at any step in the algorithm we find that all possible scattering processes are inconclusive, then the algorithm halts and completeness is not established. The minimal amount of particles required to achieve charge completeness for various groups is summarized in Table~\ref{table}.

\section{$U(1)$ Symmetry} \label{sec:U1}

Let us consider the very simplest abelian symmetry, corresponding to $G=H=U(1)$. Furthermore, we assume that the spectrum is composed of a graviton, electron, and positron whose charges are
\eq{
Q=\{0,-1,+1\}.
}{eq:U1spectrum}  
By construction, the spectrum is invariant under charge conjugation $C$.
For graviton-electron scattering we have
\eq{
0\otimes (-1) \rightarrow (-1) \lor (+1),
}{}
so the $s$- and $u$-channel charges are $-1$ and $+1$.  Since both of these are already in $Q$, no new charged particles are required, and this scattering process is inconclusive.  Meanwhile, for electron-electron scattering we have
\eq{
(-1)\otimes (-1) \rightarrow (-2) \lor 0,
}{eq:ee_scattering}
so the $s$- and $u$-channel charges are $-2$ and $0$.  Since the latter is already in $Q$, a new charged state is not guaranteed, so the scattering is again inconclusive. Said another way, while a doubly-charged state would have been required in the $s$ channel, that channel need not be activated.  If instead only the $u$ channel is present, then the dispersion relation is entirely accounted for by the graviton, which was already in the spectrum.  Thus, in this example our algorithm fails to prove completeness.
 
The situation is similarly bleak for multiple abelian factors, for example considering $G=H=U(1)^2$. Here we assume a spectrum consisting of a graviton, electron, positron, dark electron, and dark positron with charges 
\eq{
Q = \{ (0,0), (-1,0),(+1,0), (0,-1),(0,+1)\}.
}{eq:U1U1spectrum}  Obviously, any scattering involving the graviton will be inconclusive.  On the other hand, the scattering of an electron and dark electron yields
\eq{
(-1,0)\otimes (0,-1) \rightarrow (-1,-1) \lor (-1,+1).
}{eq:ede_scattering}
Both the $s$- and $u$-channel charges are new, so this pair is actually conclusive.  With charge conjugation $C$, we know that each channel grows the spectrum by the sets $\{ (-1,-1), (+1,+1)\}$ or $\{ (-1,+1) , (+1,-1)\}$.  

Concatenating either of these two sets to $Q$, we obtain two possible and distinct spectra.  However, it is easy to verify that in the next iteration of the algorithm, all scattering pairs are inconclusive.  That is, even though \Eq{eq:ede_scattering} implies the existence of new charged particles, all subsequent scattering processes can be accounted for without introducing more charged states on top of this.  This difficulty persists for any abelian symmetry.

\section{$SO(N)$ Symmetry} \label{secSON}

The plot thickens when our charges are embedded within an abelian subgroup of a nonabelian symmetry.  In this case we can actively exploit the nonabelian generators of the symmetry to transform states of a given charge into those of a different charge.  

To begin, we consider the case of a special orthogonal symmetry group $G=SO(N)$, whose Cartan subgroup is $H=U(1)^{\lfloor  N/2 \rfloor}$.  By definition, every charged particle is a simultaneous eigenstate of the generators of $H$.  The corresponding eigenvalues span the $\lfloor  N/2 \rfloor$-dimensional charge lattice of integers, 
\eq{
\Lambda  = \mathbb{Z}^{\lfloor N/2\rfloor}=\left\{ \sum_{i=1}^{\lfloor N/2\rfloor} q_i \vec{e}_i \,\middle|\, 
    q_i \in \mathbb{Z} 
     \right\},
}{Lambda_SON}
where the $\vec e_i$ are orthonormal basis vectors.  
A charge vector $\vec q \in \Lambda$ in the lattice will be denoted by
\eq{
 \vec {q} = (q_1,q_2,\ldots , q_{\lfloor  N/2 \rfloor}),
 }{}
 where each entry is an orthogonal component. 

Recall also that the center group is $Z(SO(N))=\mathbbm{1}$ or $\mathbb{Z}_2$ for odd or even $N$, respectively.  In the latter case, the central charge of a given charge vector $\vec q$ is
 \eq{
 z(\vec q) = q_1+q_2+\cdots + q_{\lfloor  N/2 \rfloor} \mod 2,
 }{}
The central charge effectively counts the parity of the number of fundamental indices of a particular representation.  We will often find it useful to classify states by their membership in the central charge sectors $z=0,1$.

On top of the structures defined in \App{groupthreview}, we will on occasion make use of the outer automorphism group $C$, which map representations to other representations.   The precise nature of $C$ will vary case by case, but it will usually correspond to some version of charge conjugation symmetry.

\subsection{$N=3$} The smallest nonabelian special orthogonal group is $G=SO(3)$.  The Cartan subgroup $H=U(1)$ furnishes a one-dimensional charge lattice defined by the integers $\Lambda = \mathbb{Z}$.  The elements of the root system, $\{ -1,+1\}$, decrement or increment the charges by unit steps. The symmetries of the root system are encoded by the Weyl group, which acts as multiplication by $-1$ and swaps the highest and lowest weights states in each representation.  The root system modulo the Weyl group yields the single simple root $\{ +1 \}$.  This structure is of course very familiar from the theory of orbital angular momentum.

Starting from the initial spectrum in \Eq{eq:U1spectrum}, we scatter as before to obtain \Eq{eq:ee_scattering} and are again confronted with the possibility of a $u$-channel state of vanishing charge.  This state is invariant under the Weyl group and functions like a ground state, which can be lowered no more by the roots.  Hence the root system and its Weyl symmetries are of little use, and it is still not possible to prove completeness for $G=SO(3)$.

\subsection{$N=4$}\label{sec:SO4}

For $G=SO(4)$ with $H=U(1)^2$, the two-dimensional charge lattice spans all pairs of integers, $\Lambda = \mathbb{Z}^2$.
 The simple roots are $ \{ (+1,+1), (+1,-1)\}$ and the Weyl group is the set of signed permutations of even signature, so for any charge vector we can use the Weyl group at will to swap its two entries or multiply by the whole vector by $-1$.

Here we will assume an initial spectrum composed of a fundamental of $SO(4)$, together with the graviton.  The charge spectrum coincides precisely with  \Eq{eq:U1U1spectrum}, so we can again scatter particles to obtain \Eq{eq:ede_scattering}.  An $s$-channel state would carry charge $(-1,-1)$, which we map to $(+1,+1)$ using the Weyl group.  Using the roots, we then lower this state to $(0,0)$.   Thus, the charge $(-1,-1)$ orbits into $\{ (-1,-1),(0,0), (+1,+1)\}$.   Analogous reasoning for the $u$ channel orbits the charge $(-1,+1)$ into $\{(-1,+1),(0,0),(+1,-1)\}$.   These $s$- and $u$-channel particles are in the self-dual and anti-self-dual two-form representations of $SO(4)$.   While these  representations are chiral, there is no inconsistency if the spectrum supports one but not the other.  Consequently, it is consistent to augment the spectrum with either the self-dual or anti-self-dual two-form.  Iterating the algorithm, one again finds that all allowed scattering processes are inconclusive, so it is not possible to prove completeness.

We can avoid this negative conclusion if we assume more symmetry.  Famously, $SO(4) \sim SO(3)^2$ admits an outer automorphism $C$ that swaps each group factor, corresponding to the orientation-reversing elements of $O(4)$.  Importantly, the action of the Weyl group and $C$ generate signed permutations of {\it any} signature, which act on a given charge vector by swapping its two entries or by multiplying any single entry by $-1$.  In what follows we will assume that $C$ is a symmetry of the dynamics.

\begin{figure}
\includegraphics[width=0.35\textwidth]{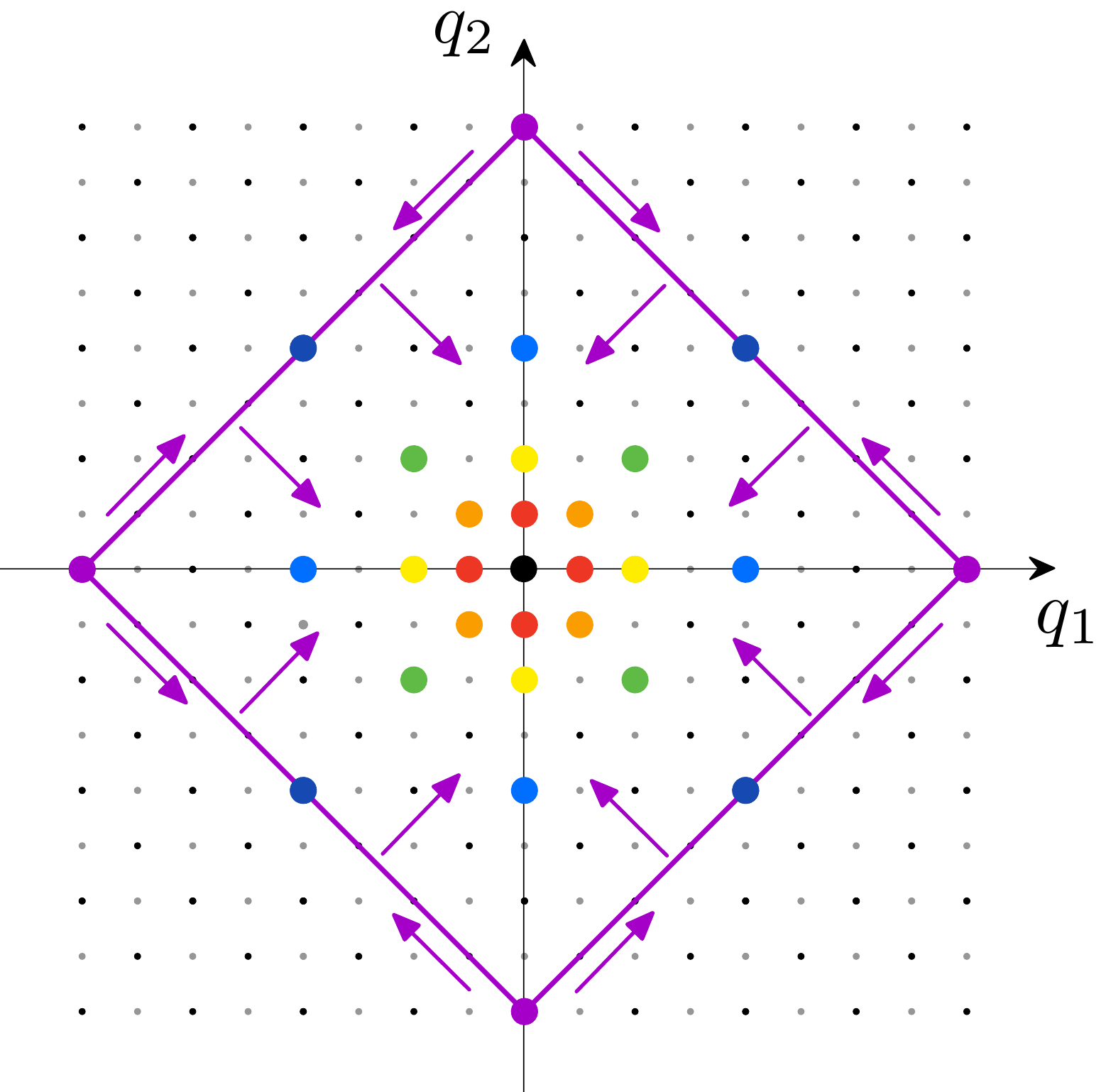}
\caption{The $SO(4)$ charge lattice, stratified according to the central charge sectors $z=0$ (black) and $z=1$ (gray).
Overlaid is the sequence of scattering processes in \Eq{eq:n_scattering}. Starting from an initial spectrum composed of the fundamental (red), we scatter in succession (orange, yellow, green, blue, indigo) to obtain a set of ultracharged states (purple).  We then apply lowering operators to generate all charges at the boundary and interior of the diamond (purple). 
}
\label{so4lattice}
\end{figure}

Armed with both $G$ and $C$, we realize that our earlier $s$- and $u$-channel sets, $\{ (-1,-1),(0,0), (+1,+1)\}$ and $\{(-1,+1),(0,0),(+1,-1)\}$, are contained within the same orbit.  Thus we can add both of these sets to the spectrum $Q$ and repeat the algorithm.

We are now equipped to prove completeness using the following sequence of scattering processes, 
\eq{
\begin{array}{r@{}c@{}l@{\;}c@{\;}r@{}c@{}l}
(1,0)      &\,\,\otimes\,\,& (0,1)      &\rightarrow& (1,1)      &\,\,\lor\,\,& (1,-1)\smallskip \\
(1,1)     &\,\,\otimes\,\,& (1,-1)     &\rightarrow& (2,0)       &\,\,\lor\,\,& (0,2)\smallskip \\
(2,0)      &\,\,\otimes\,\,& (0,2)      &\rightarrow& (2,2)      &\,\,\lor\,\,& (2,-2)\smallskip \\
(2,2)     &\,\,\otimes\,\,& (2,-2)     &\rightarrow& (4,0)       &\,\,\lor\,\,& (0,4)\smallskip \\
            &               &             &            \;\; \vdots\;\;  &              &           &  \smallskip \\
(2^n,0)    &\,\,\otimes\,\,& (0,2^n)    &\rightarrow& (2^n,2^n)  &\,\,\lor\,\,& (2^n,-2^n)\smallskip \\
(2^n,2^n) &\,\,\otimes\,\,& (2^n,-2^n) &\rightarrow& (2^{n+1},0) &\,\,\lor\,\,& (0,2^{n+1}),
\end{array}
}{eq:n_scattering}
which are depicted in \Fig{so4lattice}.
By design, we have chosen conclusive scattering processes for which the  $s$- and $u$-channel states are within the same orbit, which is to say, they are related by a signed permutation.  After $n$ iterations, the spectrum contains a set of ultracharged  corner states at $\{(2^n,0),(-2^n, 0),(0, 2^n),(0, -2^n)\}$.   These corner states reside in the $z=0$ central charge sector.
Lowering these states with the roots, we generate all states with central charge $z=0$ enclosed by these corner charges.  Clearly, in large-$n$ limit this process populates all points in the $z=0$ sector of the charge lattice.  

To derive completeness in the $z=1$ sector, we take the ultracharged corner states and scatter them against the fundamental.  For example, we find
\eq{
(2^n,0)\otimes (1,0) \rightarrow (2^n+1,0) \lor (2^n-1,0)\\
(0,2^n)\otimes (0,1) \rightarrow (0,2^n+1) \lor (0,2^n-1)
}{}
and similarly for the other corner states.  Taking the orbits of the $s$- and $u$-channel states, we obtain ultracharged corner states residing in the central charge sector $z=1$.  Again taking the limit of large $n$ and acting with lowering operators, we generate the full charge lattice for the $z=1$ sector. 
This concludes our derivation of completeness.

\subsection{$N=5$}\label{sec:SO5}
For a slightly enlarged symmetry, completeness follows even more simply.  Consider the symmetry group $G=SO(5)$, whose Cartan subgroup $H=U(1)^2$ defines the charge lattice $\Lambda = \mathbbm{Z}^2$.    Here $H$ and $\Lambda$ are the same as for $G=SO(4)$.   However, the Weyl group is enlarged to include all signed permutations of any signature.  For any charge vector, we can then swap its two entries or flip the sign of any entry.
The manipulations of the previous subsection then follow trivially, since what was an outer automorphism $C$ of $SO(4)$ is now automatically contained in $SO(5)$.  
Another way to see this fact is that the simple roots of $SO(5)$ are $\{ (+1,+1),(+1,0)\}$.  The last element can be used to toggle between the self-dual and anti-self-dual two-form representations of $SO(4)$.

\subsection{$N\geq 6$}\label{sec:SON}
We are finally ready to study the general case of $G=SO(N)$ with Cartan subgroup $H=U(1)^{\lfloor  N/2 \rfloor}$ and charge lattice $\Lambda= \mathbbm{Z}^{\lfloor  N/2 \rfloor}$.  Here we will reason by induction, assuming that completeness has already been established for all proper special orthogonal subgroups.

To begin, consider the case of odd $N$.  As usual, we assume a starting spectrum that includes the fundamental of $SO(N)$.  By extension, the spectrum carries the required fundamentals and antifundamentals of $SO(N-1)$ needed for the induction hypothesis.  
Since $SO(N-1)$ and $SO(N)$ have the exact same charge lattice and the induction hypothesis assumes completeness of the former, we have completeness of $SO(N)$ as well. 

The case of even $N$ is more involved.  The induction hypothesis assumes completeness of $SO(N-1)$, so the $SO(N)$ spectrum includes charges of the form
$ (q_1,  \ldots, q_{\frac{N}{2}-1}, 0)$,
where the nonzero entries are integer $SO(N-1)$ charges.  The last entry is chosen to be neutral.   Using the Weyl group, we can permute this zero entry to any position we like in order to construct any charge vector with one or more vanishing entries. 

Next, we construct the following pair of charges and then scatter them,
\eq{
 (q_1, \ldots, q_{\frac{N}{2}-2},0, 0)\otimes  (0, 0,\ldots, 0, q_{\frac{N}{2}-1}, q_{\frac{N}{2}}),
}{}
which result in the charges  $ (q_1,\ldots, q_{\frac{N}{2}-2},q_{\frac{N}{2}-1}, q_{\frac{N}{2}})$ or $ (q_1,  \ldots, q_{\frac{N}{2}-2},- q_{\frac{N}{2}-1}, - q_{\frac{N}{2}})$ in the $s$ or $u$ channel, respectively.  Again using the Weyl group, we apply a signed permutation of even signature to flip the sign of the final two entries, generating  $ (q_1,  \ldots, q_{\frac{N}{2}-2}, q_{\frac{N}{2}-1},  q_{\frac{N}{2}})$ to establish completeness of the spectrum.

Before moving forward, let us comment briefly on the interplay between completeness and the center of the symmetry group.   Obviously, a necessary condition for completeness is that the full spectrum contains at least one particle in all central charge sectors $z=0,1$. 
For even $N$, the center is $Z(SO(N))=\mathbb{Z}_2$ and the central charge sectors $z=0,1$ are already represented by the graviton and the fundamental.  This is a major reason why completeness arises so straightforwardly in the case of $SO(N)$: the center is tiny, and the starting spectrum already clears the low bar of containing particles with these central charges.  We will see shortly that this is not always the case when the center group is larger.

\section{$SU(N)$ Symmetry} \label{secSUN}

Let us now consider the case of $G=SU(N)$, whose Cartan subgroup is $H=U(1)^{N-1}$.   
The charge lattice is 
\eq{
\Lambda  =\left\{\vec q = \sum_{i=1}^{N-1} q_i \vec{\mu}_i \,\middle|\, 
    q_i \in \mathbb{Z} 
     \right\},
}{L_SUN}
where we emphasize that $\vec q$ is {\it not} expressed in terms of the orthonormal basis $\vec e_i$, but rather the nonorthogonal basis of fundamental weights $\vec \mu_i$, which satisfy
\eq{
\vec\mu_i = \sum_{j=i}^{N-1} \frac{i}{\sqrt{j(j+1)}} \vec e_j.
}{mu_to_e}
In particular, the components of $\vec q$ in the basis of fundamental weights are precisely the Dynkin coordinates,
 \eq{
 \vec {q} = (q_1,q_2,\ldots , q_{N-1}).
 }{}
  For this reason, one must take special care when computing dot products.
Furthermore, since $\vec e_i$ and $\vec\mu_i$ are distinct, the components of $\vec q$ are not actually the eigenvalues of the Cartan generators, but are straightforwardly related to them by a linear transformation defined by \Eq{mu_to_e}.  
  In spite of this mismatch, we will abuse nomenclature and glibly refer to $\vec q $ throughout this section as the charge vector.   Obviously, completeness in the Dynkin coordinates defined by the components of $\vec q$ will imply completeness in the bona fide charge lattice.

As described earlier, it will be useful to be able to orbit charges using the action of the symmetry group $G$ and the outer automorphisms $C$.   The former is implemented by the action of the Weyl group, which acts on the simple roots of $SU(N)$.  Explicitly, these simple roots are
\eq{
\vec r_i 
&= \sqrt{\frac{i+1}{i}}\vec e_i - \sqrt{\frac{i-1}{i}}\vec e_{i-1}\\
&=  -\vec\mu_{i-1}+2\vec\mu_{i} - \vec\mu_{i+1}, 
}{root_SUN}
where $ i = 1,2,\ldots, N-1$ and $\vec e_0 = \vec \mu_0 =\vec \mu_N=0$, with the normalization chosen so that $\vec r_i \cdot \vec \mu_j = \delta_{ij}$. In these coordinates, the Weyl group transformation reviewed in Eq.~\eqref{eq:Weyltransform} maps charge vectors via $\vec q \rightarrow \vec q - q_i \vec r_i$, which in Dynkin coordinates sends
\eq{
\begin{gathered}
 (\ldots, q_{i-1}, q_i, q_{i+1},\ldots)\\
 \downarrow \\
 (\ldots, q_{i-1}+q_i, -q_i, q_{i+1}+q_i,\ldots).
\end{gathered}
}{w_i}
This transformation flips the sign of a given entry and adds that entry to its neighbors.
On the other hand, charge conjugation $C$ reverses the order of the Dynkin coordinates via
\eq{
(q_1,q_2,\ldots , q_{N-1})\rightarrow 
( q_{N-1}, q_{N-2},\ldots ,q_1).
}{}
 For example, $C$ swaps the fundamental and antifundamental representations of $SU(N)$.

The center group $Z(SU(N)) = \mathbb{Z}_N$ will play an important role in the subsequent analysis.  The corresponding central charges of $SU(N)$ are known as $N$-ality~\cite{Georgi}, which are defined as
\eq{
z(\vec q) = q_1+2q_2+ \cdots +(N-1)q_{N-1} \mod{N}.
}{nality}
Recall that the roots only transform between states within a given central charge sector, which are labeled by $z=0,1,\ldots, N-1$. As noted previously, spanning all central charge sectors is a necessary but not sufficient condition for completeness.  Obviously, the fundamental and antifundamental, which span the sectors $z=1$ and $N-1$, are not sufficient on their own to satisfy this criterion.  This is the root of the difficulty in establishing completeness for special unitary symmetries.

\subsection{$N=3$}\label{sec:SU3}

\begin{figure}
\includegraphics[width=0.34\textwidth]{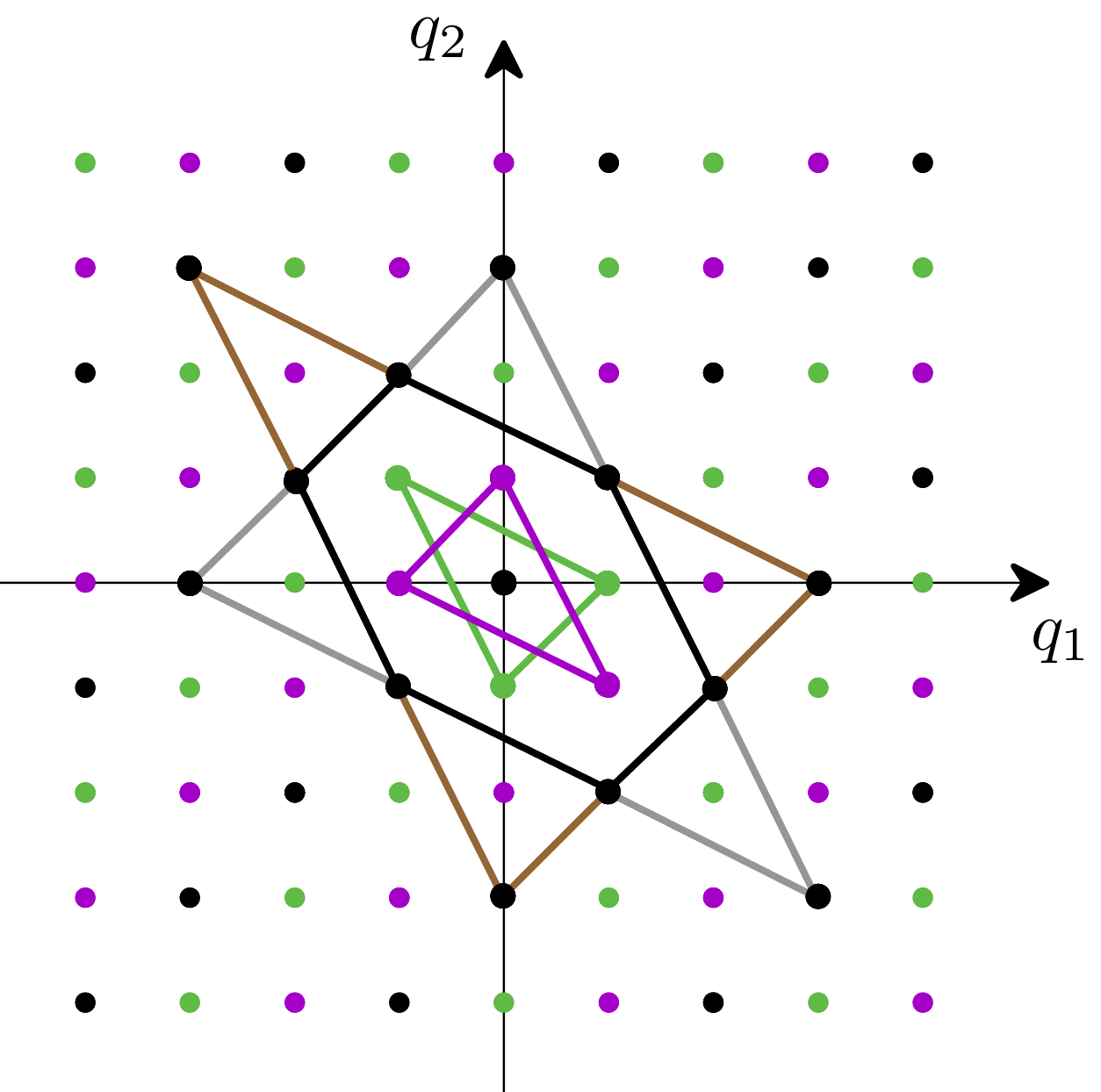}
\caption{The $SU(3)$ charge lattice, stratified according to the central charge sectors $z=0$ (black), $z=1$ (green), and $z=2$ (purple).
 The polygons circumscribe the irreducible representations $Q_{\mathbf{3}}$ (green), $Q_{ \bar{\mathbf{3}}}$ (purple), $Q_{ {\mathbf{8}}}$ (black), $Q_{ {\mathbf{10}}}$ (brown), and $Q_{ \overline{\mathbf{10}}}$ (gray).  }
\label{su3lattice}
\end{figure}

There is no reason to consider $SU(2)$ because we already considered $SO(3)$ and found that completeness could not be established using our algorithm.    We thus move on to $G = SU(3)$, whose Cartan subgroup is $H=U(1)^2$.   The corresponding charge lattice is the two-dimensional set of points defined in \Eq{L_SUN},
As shown in \Fig{su3lattice}, the charges for the singlet $\mathbf{1}$, fundamental $\mathbf{3}$, adjoint $\mathbf{8}$, and decuplet representation $\mathbf{10}$ are
\eq{
Q_{\mathbf{1}}=\,& \{(0,0) \}\\
Q_{\mathbf{3}}=\,& \{(1,0),(-1,1),(0,-1) \}\\
Q_{ {\mathbf{8}}}=\,& \left\{(1,1),(-1,2),(-2,1),(-1,-1), \right. \\
&\left.  (1,-2),(2,-1),(0,0),(0,0) \right\} \\
Q_{\mathbf{10}} = \,& \left\{(1,1),(-1,2),(-3,3),(-2,1),\right.  \\
&\left. (-1,-1),(0,-3),(1,-2),(2,-1), \right. \\
&\left. (3,0),(0,0) \right\}.
}{eq:su3spectrum}
Note that charges of the conjugate representations $Q_{\bar{\mathbf{3}}}$ and $Q_{\overline{\mathbf{10}}}$ are obtained by charge conjugation, which flips various signs in  \Eq{eq:su3spectrum}. These charge-conjugate representations are always automatically present, so we will not always explicitly enumerate them.

The $\mathbf{1}$, $\mathbf{8}$, $\mathbf{10}$, and $\overline{\mathbf{10}}$ reside in the $z=0$ charge sector, the $\mathbf{3}$ resides in $z=1$, and the $\bar{\mathbf{3}}$ resides in $z=2$. 
We now apply our usual algorithm, which is to scatter all possible states in the assumed spectrum, seeking conclusive processes where a new charged state is guaranteed in both the $s$ and $u$ channel.

For an initial spectrum comprising $Q_{\mathbf{1}}$ and $Q_{\mathbf{3}}$, we see that all possible scattering processes are inconclusive. 
In particular, for any choice of scattering between $Q_{\mathbf{3}}$ and itself or with $Q_{ \bar{\mathbf{3}}}$, there is some choice of $s$- or $u$-channel states that are still in the initial spectrum.  Hence, completeness cannot be proven.  

For the enlarged spectrum comprising $Q_{\mathbf{1}}$, $Q_{\mathbf{3}}$, and $Q_{ {\mathbf{8}}}$, we can actually do better. 
Scattering appropriately chosen representatives of $Q_{ \bar{\mathbf{3}}}$ with $Q_{ {\mathbf{8}}}$, we obtain
\eq{
(1,-1) \otimes (1,1) \rightarrow (2,0) \lor (0,-2),
}{adjsu3}
which conclusively implies the existence of a new representation.  Again orbiting the charges and applying conjugation, we learn that the spectrum must be augmented to include  new representations.
In particular, the new representations in Eq.~\eqref{adjsu3} reside in the central charge sector $z=2$, and conjugation yields the corresponding weights in the $z=1$ class.

By repeating this procedure, we accumulate a sequence of progressively larger triangles and hexagons.
At every step we orbit each charge to obtain the whole family of charges within the perimeter of the largest polygon we have reached.
This algorithm eventually covers all possible states with $z=1,2$.

As discussed in  \App{su3algs}, it is straightforward to verify that the scattering of any states within $Q_{ {\mathbf{8}}}$ is always inconclusive. As noted above, the same is true for the scattering of any state in $Q_{\mathbf{3}}$ with any state in $Q_{ \bar{ \mathbf{3}}}$.  Meanwhile, since all states within $Q_{\mathbf{3}}$ have $z=1$, scattering them will generate $z=2$ or $z=0$, while all states within $Q_{\bar {\mathbf{3}}}$ have $z=2$, so scattering them generates $z=1$ or $z=0$.
One ultimately finds that assuming an initial spectrum comprising $Q_{\mathbf{1}}, Q_{\mathbf{3}},Q_{ {\mathbf{8}}}$ conclusively generates completeness in the central charge sectors $z=1,2$ but not $z=0$.

Last but not least, let us consider an initial spectrum comprising $Q_{\mathbf{1}}$, $ Q_{\mathbf{3}}$, and $Q_{ {\mathbf{10}}}$. Pictorially, this spectrum includes the smallest triangular representation in each central charge sector. Scattering states with $Q_{ {\mathbf{10}}}$ turns out to be just enough to cover the entire $z=0$ sector via a sequence of hexagons of increasing size.   This then establishes completeness in all central charge sectors $z=0,1,2$, and more importantly, for all points in the full charge lattice.  The details of this argument are spelled out in  \App{su3algs}.

\subsection{$N\geq 4$}\label{sec:SUN}

 For the general case of special unitary symmetries, completeness requires a sufficiently large but still finite starting set of charged particles.  To see why, let us return to the accounting of central charges.  
 
A spectrum composed of the graviton, fundamental, and antifundamental spans the central charge sectors $z=0,1,N-1$.   Scattering any pair of states from this set, we see immediately that either the $s$ or the $u$ channel will contain a state in $z=0,1,N-1$.  This implies that we can never conclusively guarantee a new charged particle with central charge $z=2, 3, \ldots, N-2$.  

Now imagine that we instead augment our starting spectrum to include states residing in $z=0,1,2, N-2,N-1$.  By the same logic as before, by scattering these states we can never conclusively guarantee a new charged particle  with $z=3, 4, \ldots, N-3$, and so on.  This reasoning implies that completeness will always require an initial spectrum of states with representatives in all central charge sectors $z=0,1,\ldots, N-1$.  Any initial spectrum that fails this criterion will not ensure full charge completeness across all sectors.  

Remarkably, the existence of a representative in all central charge sectors $z=0,1,\ldots, N-1$ is a sufficient condition for completeness.   This is very much not obvious.
To see why this holds, let us start with the example of deriving completeness of the $z=0$ sector assuming that the spectrum contains the adjoint representation.
 Consider the following scattering process involving states within the adjoint representation,
 \eq{
    \vec r_1 \otimes \vec r_3 \rightarrow (\vec r_1+\vec r_3) \lor (\vec r_1-\vec r_3) .
    }{}
Here we have used that the charges of the adjoint are themselves simple roots.
Importantly, both the $s$- and $u$-channel exchanges are related by action of the Weyl group.  In particular, let us we define $w_i$ to be the Weyl transformation in \Eq{w_i}.  Then we see that  $(\vec r_1+ \vec r_3)$ and $(\vec r_1- \vec r_3)$ are related by the action of $w_3$, so this scattering is conclusive.  We then scatter this pair of charges with each other to obtain
    \eq{
    (\vec r_1+\vec r_3) \otimes (\vec r_1-\vec r_3) \rightarrow (2 \vec r_1) \lor (2 \vec r_3),
}{}
where the $u$- and $s$-channels are related by a sequential composition of Weyl transformations, $w_2 \circ w_1 \circ w_3 \circ w_2$.

After $n$ iterations back and forth between these two classes of scattering processes, we eventually obtain all charges of the form  $2^n \,\vec r_1$ and $2^n \,\vec r_3$, which are charges of the adjoint representation multiplied by $2^n$.
By orbiting these two charges via the weight strings reviewed in \App{groupthreview}, we obtain all of the adjoint charges, multiplied by $2^n$.   These sets of charges are concentric polytopes, each twice the size of the previous one, and they reside within an ever larger set of representations.  We then use lowering operators of the symmetry group to generate all states enclosed by these charges in the $z=0$ central charge sector.   At large $n $ we ultimately recover the entire $z=0$ sector.
The procedure here is obviously closely analogous to the one used to derive completeness in $SO(4)$ in \Fig{so4lattice}.
Crucially, we note that the algorithm just outlined requires the existence of $\vec r_3$. Since $SU(N)$ has $N-1$ roots, as defined in \Eq{root_SUN}, this construction highlights why $N=3$ is not enough for completeness. 

With completeness in the $z=0$ sector established, it is now straightforward to show that the inclusion of even a single charge with $z=k$ or $z=N-k$ is sufficient to ensure completeness in the entire $z=k$ and $z=N-k$ sectors. This sector-wise charge completeness is achieved as follows. First, we pick any charge in the $z=0$ sector of the form $n N \vec \mu_k$.  Next, given any charge in the class $z=k$, we are guaranteed the existence of the fundamental weight $\vec\mu_k$, since as we have seen we can obtain this charge vector from our starting charge via Weyl transformations and weight string relations. We then scatter them to obtain
   \eq{
    (n N \vec \mu_k) \otimes \vec \mu_k \rightarrow ((n N+1) \vec \mu_k) \lor  ((n N-1) \vec \mu_k) .
}{}
If the $s$-channel exchange is activated, then we obtain the charge $(n N+1) \vec \mu_k$, which is manifestly in sector $z=k$. Furthermore, the orbit of this state generates a polytope that is simply a rescaled version of $\vec \mu_k$ with vertices expanded outwards to inflate the polytope along all directions.  Orbiting these charges into the interior, we then obtain all $z=k$ states inside the polytope, and by charge conjugation $C$ we obtain the same in the $z=N-k$ sector.  In the large-$n$ limit, this construction implies completeness in the charge sectors $z=k, N-k$. 
Alternatively, if we instead have the particle in the $u$ channel with charge $(n N-1) \vec \mu_k$, this state is in sector $z=N-k$.  In this case, the same conclusion holds, only with the roles of $N-k$ and $k$ swapped. 
As advertised, this scattering process establishes that the central charge sectors $z=k, N-k$ are complete, provided we already have completeness in the $z=0$ sector and that we have a single charge with either $z=k$ or $z=N-k$. We again emphasize that in the proof we used a fundamental weight of class $k$.   If we started with any other representative, then the weights of the fundamental would automatically be part of the spectrum by the properties of representations discussed in \Sec{sec:initialization}.

\section{$Spin(N)$ Symmetry}\label{sec:SpinN}

The spin group $Spin(N)$ is the well-known universal cover of the special orthogonal group $SO(N)$.  By definition, these groups share the same roots and hence the same Weyl group.  They only differ in their charge lattices, with  the weight lattice of $Spin(N)$ given by 
\eq{
\Lambda &= \Lambda_0 \cup \Lambda_{1/2}\\
\Lambda_0  &= \mathbb{Z}^{\lfloor N/2\rfloor}=\left\{ \vec q=\sum_{i=1}^{\lfloor N/2\rfloor} q_i \vec{e}_i \,\middle|\, 
    q_i \in \mathbb{Z} 
     \right\}\\
\Lambda_{1/2}  &= (\mathbb{Z}{+}\tfrac12)^{\lfloor N/2\rfloor}=\left\{\vec q\,{=}\!\! \sum_{i=1}^{\lfloor N/2\rfloor} q_i \vec{e}_i \,\middle|\, 
    q_i \,{\in}\, \mathbb{Z} {+}\tfrac12
     \right\},\!
}{}
where $\Lambda_0$ is just the integer lattice of $SO(N)$ and $\Lambda_{1/2}$ is the same lattice shifted by a half-integer in every direction.    The latter correspond to the spinor representations of $Spin(N)$ that are not in  $SO(N)$.  The group $Spin(N)$ admits the $N$-dimensional vector representation, together with the  
$2^{\lfloor (N-1)/2\rfloor}$-dimensional
spin representations of multiplicity one and two for $N$ odd and even respectively.  

As per our earlier arguments, the existence of a vector will guarantee completeness in the $\Lambda_0$ charge lattice, which corresponds to $SO(N)$.
Obviously, to ensure completeness in $\Lambda_{1/2}$ as well, we also have to include spinor representations of $Spin(N)$ in the spectrum.  

For these reasons, we will assume that our spectrum contains the vector $\vec e_i$ and the spinors $\frac{1}{2}\sum_i \sigma_i \vec e_i$ for all sign choices $\sigma_i \in \{\pm 1\}$.  For even $N$, the cases where the $\sigma_i$ sum to even or odd constitute the two different spinors.
Equipped with these states, it is then straightforward to generate the full charge lattice. 

We take the ultracharged corner states constructed in the previous section and scatter them against the spinors. In the case of $Spin(4)$, this scattering yields
\eq{
\begin{array}{r@{}c@{}l@{\;}c@{\;}r@{}c@{}l}
(2^n,0)    &\,\,\otimes\,\,& (\tfrac12,\tfrac12)     &\rightarrow&  (2^n+\tfrac12,\tfrac12)     &\,\,\lor\,\,&  (2^n-\tfrac12,-\tfrac12) \smallskip \\
(0,2^n)     &\,\,\otimes\,\,& (\tfrac12,\tfrac12)   &\rightarrow&  (\tfrac12,2^n+\tfrac12)       &\,\,\lor\,\,& (-\tfrac12,2^n-\tfrac12).
\end{array}
}{}
For $Spin(N)$, we would use the corresponding corners of the $\lfloor N/2\rfloor$-dimensional orthant.
These scattering processes generate ultracharged states in the half-integer lattice, which we can then orbit inwards to generate all enclosed charges within the same central charge sector.  Taking the large-$n$ limit, we then derive completeness of the full spectrum across the integers and half-integers.  In conclusion, $Spin(N)$ is complete, provided the spectrum includes the vector and spinor representations.

\section{ $Sp(N)$ Symmetry}\label{sec:SpN}

Consider the compact symplectic group $Sp(N)$, which defines the set of linear transformations that leave invariant the symplectic form in even $N$ dimensions.  Here, completeness is an immediate consequence of our earlier results.
To see why, note that $Sp(N)$ has the very same charge lattice as $SO(N)$, namely $\Lambda=\mathbb{Z}^{N/2}$.
In particular, the vector representation of $Sp(N)$ has the same charges as in $SO(N)$.
Furthermore, the root system of $Sp(N)$ contains that of $SO(N)$ as a subset, so the same is true for their respective Weyl groups.
This implies that with respect to orbits, $Sp(N)$ is availed of strictly more operations to generate new charges from old charges. Altogether, these relations imply that all of the scattering and orbit operations we performed for $SO(N)$ are equally applicable for $Sp(N)$.
This establishes completeness of the symplectic group $Sp(N)$.

\section{$E_8$ Symmetry} \label{sec:E8}

We can make an analogous argument for $E_8$, the largest exceptional simple Lie group.
Unlike $SO(N)$ and $SU(N)$, the smallest nontrivial representation of $E_8$ is the adjoint $\bf 248$.
We will therefore assume an initial spectrum $Q$ comprising the graviton and the adjoint.
The charge lattice of $E_8$ is famously even and self-dual, 
\eq{
\Lambda &= \Lambda_0 \cup \Lambda_{1/2}\\
\Lambda_0  &= \left\{ \vec q=\sum_{i=1}^{8} q_i \vec{e}_i \;\middle|\;
    q_i \in \mathbb{Z}, \;\; \sum_{i=1}^8 q_i \in 2\mathbb{Z}
     \right\}\\
\Lambda_{1/2}  &= \left\{ \vec q=\sum_{i=1}^{8} q_i \vec{e}_i \;\middle|\;
    q_i \in \mathbb{Z} +\tfrac12 , \;\; \sum_{i=1}^8 q_i \in 2\mathbb{Z}
     \right\},
     }{eq:LE8}
     and is equivalent to its root lattice. 
In particular, the root system comprises $R_0$, the set of all $\pm \vec e_i \pm \vec e_j$ with independent signs, along with $R_{1/2}$, the set of all points in $\Lambda_{1/2}$ with all $|q_i| = 1/2$. Together with the point at the origin, $R_0$ and $R_{1/2}$ are the adjoint weights of $E_8$.

We immediately recognize $\Lambda_0$ and $R_0$ as precisely the $z=0$ sector weight lattice and root system of $SO(16)$.
For $SO(N)$, the full $z=0$ sector is generated by our scattering algorithm if we start with the adjoint rather than the vector, as is clear beginning with the second line of Eq.~\eqref{eq:n_scattering}, followed by Weyl symmetry and lowering operators as described in that section.
In the exact same way, for $E_8$ we are forced to augment $Q$ to all of $\Lambda_0$ by running the scattering algorithm starting from the adjoint weights in $R_0$.

In particular, we note that the point $2^n (\vec e_1 + \vec e_2)$ is in $\Lambda_0$ and hence in $Q$.
Using the Weyl orbit of this weight, in parallel with our earlier constructions we obtain a polytope composed of the $E_8$ adjoint weights rescaled by $2^n$.
Applying the results reviewed in \App{groupthreview}, along with the fact that the center of $E_8$ is trivial, we conclude that all points in the full lattice $\Lambda$ contained within this polytope must also be in representations described by the polytope,  so these weights must be included in $Q$ as well.
As $n$ tends to infinity, we find that $Q = \Lambda$ and hence conclude that charge completeness of $E_8$ follows from the presence of the adjoint and the graviton.  

\vspace{-1mm}   

\section{Grand Unified Theories} \label{sec:GUT}

\vspace{-0.8mm}

The above analysis has direct implications for the charge completeness of GUTs. As we will see below, in these theories, the field content of the standard model actually  {\it ensures} charge completeness.

\vspace{-5mm} 

\subsection{$G = SU(5)$}\label{sec:GUT5}

\vspace{-0.8mm}

The minimal GUT is the Georgi-Glashow model~\cite{Georgi:1974sy}, whose gauge group is $G=SU(5)$.  The gauge bosons transform in the $\mathbf{24}$, while the Higgs fields that break the electroweak and grand unified symmetries reside in the  $\mathbf{5}$ and $\mathbf{24}$.
The quarks and leptons of the standard model fit snugly within the $\bar{\mathbf{5}}$ and $ \mathbf{10}$.

As we saw in \Sec{sec:SUN}, the existence of a particle in each central charge sector in the initial spectrum is enough to guarantee full charge completeness.  In this case the center group is  $Z(SU(5))=\mathbb{Z}_5$.    The sectors $z=0,1,2,3,4$ are accounted for by  $ \mathbf{24}, \mathbf{5},\mathbf{10},\overline{\mathbf{10}}, \bar{\mathbf{5}}$, respectively. 
We thus have completeness of all classes $z=0,1,2,3,4$ in the $SU(5)$ GUT. 

Curiously, there is an intriguing correlation between the requirements of completeness and those needed for viable phenomenology.  In particular, according to our arguments about central charges, any strict subset of the representations $\mathbf{5}, \mathbf{10},\mathbf{24}$ and their conjugates are {\it insufficient} to imply completeness.  Said another way, a theory with $SU(5)$ symmetry that is missing any of the quarks, leptons, or gauge bosons of the standard model would not furnish enough charged states to ensure completeness using our logic.

\vspace{-2mm}

\subsection{$G = Spin(10)$}\label{sec:GUT10}

\vspace{-0.8mm}

An alternative scheme that naturally incorporates all fermions into a single representation is  $G=SO(10)$ grand unification~\cite{Fritzsch:1974nn}, or more precisely its double cover $G=Spin(10)$.  The gauge bosons reside in the $\mathbf{45}$, while the $\mathbf{16}$ includes the standard model quarks and leptons plus the right-handed neutrino.  The Higgs resides in the $\mathbf{10}$, while the fields responsible for breaking the grand unified symmetry can involve even more representations, depending on the model.

The center of the gauge group is $Z(Spin(10))=\mathbb{Z}_4$, so to guarantee full charge completeness we need a spectrum that includes states in the equivalence classes $z=0,1,2,3$.
As we saw in our $SO(N)$ construction, the $z=0$ class can be generated by producing the adjoint $\mathbf{45}$ by scattering the fundamental $\mathbf{10}$. 
The required charges $z=1,2,3$ are thus provided by  $\mathbf{16},\mathbf{10},\overline{\mathbf{16}}$, respectively, which are all the ingredients that embed the matter content of the standard model. Remarkably, we see that any strict subset of these representations, which would omit some standard model matter content, does not guarantee full charge completeness.

\vspace{-1.5mm}

\section{Discussion} \label{sec:discussion}

\vspace{-1.5mm}

We have shown that the spectrum of charges is complete across a range of theories under mild assumptions.  The cornerstone of our analysis is the dispersion relation for gravitational scattering amplitudes in \Eq{DR}, first proposed in Ref.~\cite{Hillman:2024ouy}. This remarkable formula directly relates the scattering contribution from graviton exchange to a sum over exchanges in the $s$ and $u$ channel.  As we have emphasized, the assumption of a tree-level ultraviolet completion of gravity is an important criterion for our analysis.  This condition not only zeros out the boundary term in \Eq{DR}, but also dictates that the unitarizing degrees of freedom are single-particle states.
Under these conditions, we have arrived at a surprisingly strong conclusion: in a litany of cases, given the existence of a symmetry $G$ and a handful of charged seed states, the sheer presence of gravity directly mandates the existence of single-particle states charged under {\it all possible} Cartan charges of $G$.  This follows purely as a consequence of the self-consistency of scattering amplitudes.
If the standard model resides in an ultraviolet completion with this symmetry $G$, then these assumptions require the existence of new particles with doubly, triply, etc. charged excitations of the leptons and quarks.

Our proof exploits the sequential scattering of particles and the action of the symmetry group $G$ itself to generate the full spectrum of charges~\footnote{
Throughout this work, we will assume that $G$ is a single group factor.  For a product group,  $G = \otimes_i G_i$, completeness of each $G_i$ factor implies completeness of $G$ under certain conditions. In particular, consider the case in which a starting spectrum $Q_i$ is sufficient to prove completeness for $G_i$. First, we require an initial spectrum $Q$ for $G$ composed of all states with charges in $Q_i$ under $G_i$ and neutral under all others, for each $i$.  Second, we assume that each $G_i$ has no outer automorphisms, which would induce a relative alignment between charges under different group factors.  If these conditions are met, completeness of the product group $G$ is guaranteed.}. For an abelian symmetry $G$, we are unable to prove completeness in any  form.  However, for certain choices of nonabelian $G$, we show that charge completeness is mathematically required as long as the spectrum contains at least some finite handful of charged states, usually taken to be the fundamental.   In our context, charge completeness is the property that the full charge lattice of the Cartan subgroup $H$ is populated by single-particle states.

Our results suggest a number of avenues for future work.  First and foremost is the question of whether it is possible to derive more general forms of completeness.  In particular, given that charge completeness arises relatively straightforwardly, it is natural to ask: are all {\it irreducible representations} of a nonabelian symmetry $G$ required to be in the spectrum? 
Of course, charge completeness necessitates the presence of an infinite collection of representations of arbitrarily high weight.
  We have initiated a partial investigation into the question of representation completeness, yielding primarily negative results.  
We elaborate on our various attempts in App.~\ref{app:reprcompl}, based on the strategy of Ref.~\cite{Hillman:2024ouy}.  We present evidence that in the cases of  $G=SO(3), SO(4)$, representation completeness cannot be proven using this methodology.  
More generally, given that the entirety of our analysis has focused on four-point scattering, it is worth examining whether higher-point processes might afford more leverage.  In particular, recent work has shown that positivity constraints on higher-point scattering are exceedingly stringent~\cite{Cheung:2025nhw}.

A second question relates to the precise nature of the symmetry $G$.  As discussed earlier, the conclusions of the present work are independent of whether $G$ is gauged or global.  
The latter famously runs afoul of the expectation that quantum gravity forbids exact global symmetries.  This does not detract from our logic, however, since we are not claiming that a global symmetry is required.  Indeed it would be interesting to study the case where the symmetry is explicitly broken.  Conversely, if the symmetry is gauged then any logical implications of the dispersion relation could be relevant to the weak gravity conjecture~\cite{Arkani-Hamed:2006emk}.  

Furthermore, while we have assumed throughout that $G$ is internal, another interesting possibility is that $G$ could be a spacetime symmetry, for example the Poincar\'e group.  Since spacetime symmetries also imply conservation laws, it is natural to speculate on the completeness of spacetime charges such as physical spins.  Spinning states of this kind are precisely what is needed to explicitly break the higher-form symmetries of gravity~\cite{Cheung:2024ypq}.  Of course, our assumptions already imply spin completeness: reproducing the $1/t$ pole on the left-hand side of \Eq{DR} requires an infinite tower of spins on the right-hand side, since each partial wave is a polynomial in $t$~\footnote{This can also be derived using the null constraints introduced in Ref.~\cite{Caron-Huot:2020cmc}.}.  A corollary of this fact is that the existence of particle of a given charge implies the existence of an infinite tower of higher-spin cousins with that same charge. 

A third line of inquiry concerns whether our results could be strengthened or enriched by relaxing existing assumptions or adding new ones.  Clearly, our strongest assumption is that the relevant dynamics are at tree level, so it would be worthwhile to understand the effects of loops. At a technical level, loops would allow for multi-particle states in the dispersion relation in \Eq{DR}, which naively ensures completeness trivially, starting from some initial seed of charged states.    Another central assumption is the presence of the graviton, so it would be interesting to examine how important gravity truly is for our conclusions. What we fundamentally require is a particle that couples universally, so that it shows up in all of our dispersion relations, and whose amplitudes exhibit sufficiently soft high-energy behavior. Studying other setups that share these properties could be illuminating. 

Alternatively, since our analysis is relatively conservative it is also reasonable to include stronger assumptions.  For example, our arguments make no use of unitarity, let alone detailed kinematic information other than the dispersion relation itself.
Unitarity is obviously a very weak assumption, but it would likely provide additional mileage. 
 Furthermore, if we could somehow know for a fact that both the $s$ and $u$ channels of \Eq{DR} are always activated, then much stronger claims could be proven. 
We discuss this  possibility briefly in App.~\ref{app:channels}. 

Last but not least, our results suggest that there is untapped potential in applying the methods of the scattering bootstrap to other conjectures in quantum gravity from the bottom up.  The modern amplitudes program and the bootstrap approach have recently demonstrated notable utility in addressing problems in quantum gravity, from the uniqueness of string theory to, in this work, questions in the swampland program such as the completeness hypothesis. The possibility of applying these techniques to other questions seems especially likely given the close relationships linking the completeness hypothesis to the weak gravity conjecture and the absence of global symmetries.
The intersection of top-down and bottom-up approaches to quantum gravity, in applying the amplitudes bootstrap to gravitational scattering, offers a compelling program for future study, instantiating a powerful new set of tools for addressing these fundamental questions.
 
\medskip

\noindent {\it Acknowledgments:} 
We thank Nima Arkani-Hamed, Jos\'e Calder\'on-Infante, Aaron Hillman, Yu-tin Huang, and Julio Parra-Martinez for useful discussions and comments on the paper.
F.C., C.C.,~and M.T.~are supported by the Department of Energy (Grant No.~DE-SC0011632), the Walter Burke Institute for Theoretical Physics, and the Leinweber Forum for Theoretical Physics. 
G.N.R. is supported by the James Arthur Postdoctoral Fellowship at New York University. F.S.~is
supported by the research grants 2021-SGR-00649, PID2023-146686NB-C31, and funding from the European Union NextGenerationEU (PRTR-C17.I1).

\bibliographystyle{utphys-modified}
\bibliography{completeness,completeness_arxivNotes}


\appendix

\setcounter{equation}{0}
\renewcommand{\theequation}{A\arabic{equation}}

\numberwithin{equation}{section}

\section{Group Theory Review}\label{groupthreview}

In this appendix, we review some of the group theoretic structures needed to establish charge completeness.
Consider a finite, semisimple, compact Lie group $G$ with an associated Lie algebra  $\mathfrak{g}$.  
The maximal commuting subalgebra of $\mathfrak{g}$ is the Cartan subalgebra $\mathfrak{h}$, which generates the Cartan subgroup $H$.   The elements of $\mathfrak{h}$ are the generators $H_i$.  The eigenvalues of these generators are defined by $H_i | \vec q \rangle = q_i |\vec q \rangle$, where $q_i$ are the weights, or charges.

Particularly important to our discussion are the root generators, which are operators $E_{\vec r_i}$ satisfying $[H_i,E_{\vec r_i}]=\vec r_i E_{\vec r_i}$. The root generators are each labeled by a vector $\vec r_i$, known as the root, which all together form the root system.  These roots move between weights of the lattice according to $E_{\vec r_i} |\vec q \rangle \propto |\vec q + \vec r_i \rangle$. The roots themselves are also weights, corresponding to the adjoint representation.
The group $G$ may have a center $Z(G)$, defining the set of elements that commute with every element of $G$. 
A nontrivial center defines charges under the center symmetry for each representation, and we say that representations with the same charges under $Z(G)$ are in the same central charge sector.
A pair of weights reside in the same central charge sector if and only if they can be connected by an integer linear combination of roots. Note that the fundamental weights $\vec \mu_i$ are defined to be orthogonal to the roots, so $2\, \vec \mu_i \cdot \vec r_j/| r_j |^2 = \delta_{ij}$.
The fundamental representations are the irreducible representations whose maximal weight is a fundamental weight.

For a given charge $\vec q$, it is natural to classify each root $\vec r$ according to whether it points ``towards the origin'' or ``away from the origin,'' corresponding to $\vec q \cdot \vec r <0$ or $\vec q \cdot \vec r >0$, respectively.  This classification defines generalized raising and lowering operators that increase or decrease the magnitude of the charge vector.   Crucially, if $\vec q \cdot \vec r < 0$, then the weight string formula~\cite{Humphreys} says that both $\vec q$ and its lowered cousin $\vec q+ \vec r$ reside in the same representation.
Concretely, the set of states $\vec q + k \vec r$ are all in the same representation for all integers $k = -m, -m+1,0,1,\ldots,n$, where $m$ and $n$ are nonnegative integers satisfying 
\eq{
m-n = \frac{2 \vec q \cdot \vec r}{\vec r \cdot \vec r}.}{eq:WSL}
Hence, for $\vec q \cdot \vec r < 0$, we have $n>m\geq 0$, and so $\vec q + \vec r$ is in the same representation as $\vec q$. Similarly, of course if $\vec q \cdot \vec r > 0$, then $m>n\geq 0$, and so $\vec q - \vec r$ is in the same representation as $\vec q$.

A second ingredient is that the root system exhibits isometries parameterized by the Weyl group.
By definition, the latter reflects any given charge vector $\vec q$ through the plane orthogonal to any root $\vec r$, 
\eq{ \vec q \rightarrow  \vec q^{\,\prime} = \vec q - \frac{2(\vec q \cdot \vec r)}{(\vec r \cdot \vec r)} \vec r.
}{eq:Weyltransform}
For $SO(N)$, these Weyl transformations act on $\vec q$ as a signed permutation of any signature for odd $N$ and of even signature for even $N$.  For $SU(N)$, the action of the Weyl group is more complicated.

Importantly, any two points $\vec q$ and $\vec q^{\,\prime}$ related by a Weyl transformation define a line segment,
\eq{
\vec q(\lambda) = \lambda \vec q + (1-\lambda) \vec q^{\,\prime} = \vec q - 2(1-\lambda)\frac{\vec q\cdot \vec r}{\vec r\cdot \vec r} \vec r,
}{}
where $\lambda \in [0,1]$.
Going from $\vec q$ to $\vec q^{\,\prime}$ moves $\vec q$ in the direction of $-\vec r$ for $\vec q \cdot \vec r > 0$ and in the direction of $+\vec r$ for $\vec q \cdot \vec r < 0$.
In either case, due to the weight string formula in Eq.~\eqref{eq:WSL}, all charges on this line segment connecting $\vec q$ and $\vec q^{\, \prime}$ must lie in the same representation.  This mathematical fact allows us to ``connect the dots'' between charges in order to orbit charges circumscribing the boundaries of the charge sets of representations.
In general, the smallest irreducible representation containing a given weight $\vec{q}$ also contains all the weights in the same central charge sector as $\vec{q}$ that reside inside the convex hull defined by this perimeter~\cite{Hall2015}.

\section{$SU(3)$ Symmetry}\label{su3algs}

In the following discussion, we elaborate on the details behind our completeness results for $G=SU(3)$. We will consider the cases in which the initial spectrum is $Q_{\mathbf{1}}, Q_{\mathbf{3}},Q_{ {\mathbf{8}}}$ and $Q_{\mathbf{1}}, Q_{\mathbf{3}},Q_{ {\mathbf{10}}}$, where completeness can be derived for $z=1,2$ and $z=0,1,2$, respectively. In both cases, after a few scattering processes, we will identify a clear recursive pattern that guarantees completeness in the claimed central charge sectors.

 \subsection{ $Q_{\mathbf{1}}, Q_{\mathbf{3}},Q_{ {\mathbf{8}}}$ Spectrum}

\begin{figure}
\includegraphics[width=0.46\textwidth]{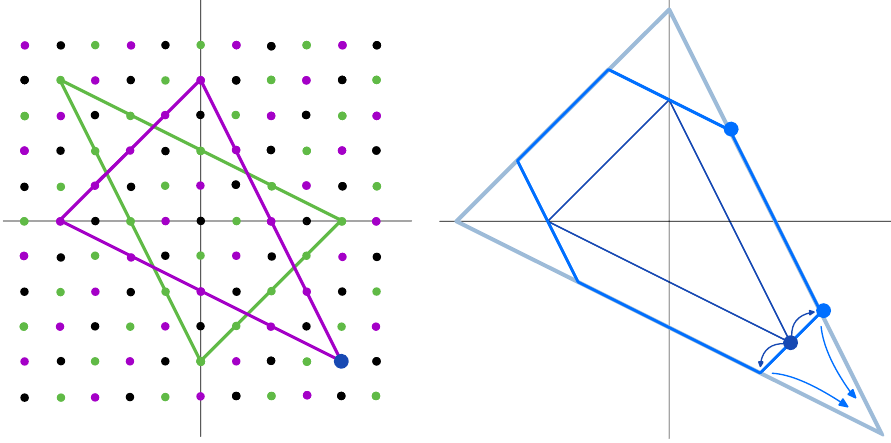}
\caption{The $SU(3)$ charge lattice. Left: The seed of the iteration. Right: Sequence of polygons allowing iteration, namely, the dark blue triangle $T_n$, blue hexagon $H_n$, and light blue triangle $T_{n+3}$. Relevant points for scattering are denoted with circles.}
\label{su3alg1}
\end{figure}

Starting from $Q_{\mathbf{1}}, Q_{\mathbf{3}},Q_{ {\mathbf{8}}}$, we consider the following three scattering processes,
\eq{
(1,-1) \otimes (1,1) &\rightarrow (2,0) \lor (0,-2) \\
(2,-2) \otimes (1,1) &\rightarrow (3,-1) \lor (1,-3) \\
(3,-2) \otimes (1,2) &\rightarrow (4,0) \lor (2,-4),
}{adjsu3alg1}
where we use the Weyl transformation in \Eq{w_i} and charge conjugation after each process.  These scattering processes yield the charges comprising the two triangles in the left panel of \Fig{su3alg1}, which correspond to the new representations  $Q_{{\mathbf{15}}}$ and  $Q_{\overline{\mathbf{15}}}$. In the last line we are guaranteed the existence of a state of charge $(4,0)$ or $(2,-4)$.  If the state $(2,-4)$ is activated,  then since this state is a representative of the $z=0$ sector, this implies completeness in $z=0,1,2$ and we claim victory.  To be conservative, we instead assume that the other state $(4,0)$ is activated.

Now we are ready to state the iterative procedure, starting from the seed representation $Q_{\overline{\mathbf{15}}}$.  The charges at the corners of $Q_{\overline{\mathbf{15}}}$ reside at $\{(4,-4),(0,4),(-4,0)\}$.  For later convenience, let us define the sequence of triangles $T_n = \{(n,-n),(0,n),(-n,0)\}$, where $T_4$ are the corners of $Q_{\overline{\mathbf{15}}}$.
Note from Eq.~\eqref{w_i} that all three corners of $T_n$ are contained within the same Weyl orbit. We then apply the following algorithm:

\begin{itemize}[
    labelindent=0pt,       
    leftmargin=0.2cm,        
    rightmargin=0.2cm,       
    labelsep=0.5em,        
    align=left             
]

\item[] {\it i}) Assuming that we already have the corner charges in $T_n$, consider the scattering process
\eq{
\begin{aligned}
(n,-n) &\otimes (1,1)\\
 &\hspace{1pt}\downarrow \\
(n+1,-n+1) &\lor (n-1,-n-1).
\end{aligned}
}{}
The $s$- and $u$-channel states above map are related via the Weyl transformation $w_1 \circ w_2 \circ w_1$, so the process is conclusive and we append both weights to $Q$. We use the Weyl group and charge conjugation to construct the full hexagon of charges $H_n$ from $T_n$.  This is depicted geometrically as a blow-up operation in the right panel of \Fig{su3alg1}. Explicitly, one of the Weyl transformations $w_2$ sends $(q_1,q_2) \rightarrow (q_1+q_2,-q_2)$, which connects the two blue points in the figure, $(n+1,-n+1)$ and $(2,n-1)$.
\item[] {\it ii}) Take charges in the hexagon and scatter them, 
\eq{
\begin{aligned}
(n+1,-n+1) &\otimes (2,n-1) \\
&\hspace{1pt} \downarrow \\
 (n+3,0) &\lor (n-1,-2n+2).
\end{aligned}
}{}
Immediately, we see that the state $ (n-1,-2n+2)$ is in the $z=0$ sector, so its presence would ensure completeness in $z=0,1,2$.  Again, we choose the more conservative option and assume that the activated channel is $(n+3,0)$.  Applying the Weyl group and charge conjugation, we obtain the corners of the new triangle $T_{n+3}$. This result is shown geometrically by the outermost triangle in the right panel of \Fig{su3alg1}.
\end{itemize} 
We then repeat these steps ad infinitum to cover the $z=1$ and $z=2$ sectors fully via a sequence of alternating triangles and hexagons.

  \subsection{ $Q_{\mathbf{1}}, Q_{\mathbf{3}},Q_{ {\mathbf{10}}}$ Spectrum}

\begin{figure}[h]
\includegraphics[width=0.46\textwidth]{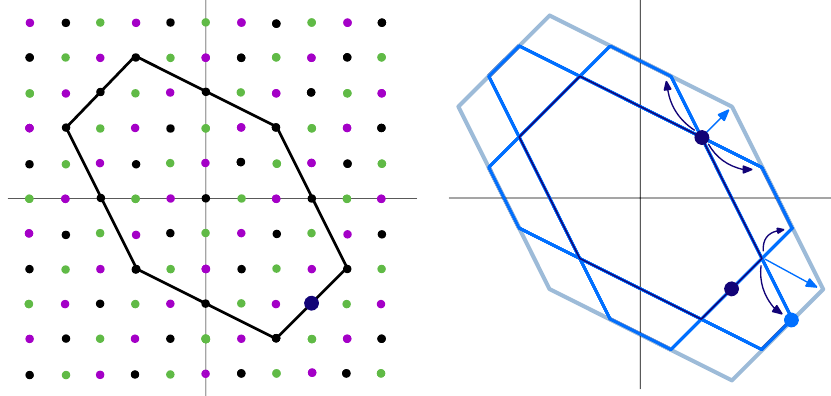}
\caption{The $SU(3)$ charge lattice.  Left: starting hexagon with the starting charge $P_0$ (dark blue). Right: sequence of hexagons (dark blue, blue, light blue), highlighting the relevant points $P_n$, (dark blue), $D_n$ (dark blue), and $P_{n+1}$ (blue).}
\label{su3alg2}
\end{figure}

For $Q_{\mathbf{1}}, Q_{\mathbf{3}},Q_{ {\mathbf{10}}}$, the initial spectrum includes the brown and gray triangles in \Fig{su3lattice}.
As before, we consider a wisely chosen scattering process to position ourselves for a recursive procedure,
\eq{
(3,-3) \otimes (1,1) \rightarrow (4,-2) \lor (2,-4)
}{eq:22}
 Crucially, via Weyl symmetry, the input of the triangles in \Fig{su3lattice} allows us to take the full orbit of the right side of Eq.~\eqref{eq:22} to get all the points in the black hexagon depicted in \Fig{su3alg2}, in particular the new charge $(2,2)$. This would not have worked assuming only the adjoint.

Next, we define the specific points $P_n = (3+2n,-3-n)$ and $D_n = (2+n,2+n)$, denoted by the large dots in the right panel of \Fig{su3alg2}. Our  starting points will be $P_0$ and $D_0$. Now iterate as follows.

\begin{itemize}[
    labelindent=0pt,       
    leftmargin=0.2cm,        
    rightmargin=0.2cm,       
    labelsep=0.5em,        
    align=left             
]

\item[] {\it i}) At step $n$, consider the scattering process,
\eq{
P_n \otimes D_n \rightarrow (5+3n,-1) \lor (1+n,-5-2n).
}{}
Note that the $s$-channel charge resides in a larger representation than the $u$-channel charge, since $|5+3n| \geq |5+2n| > |1+n|$.  To be conservative, we assume the smaller representation, which corresponds to the latter choice of  $(1+n,-5-2n)$.
This operation is geometrically understood in \Fig{su3alg2} as blowing up each edge of the dark blue hexagon to get the blue one.
\item[] {\it ii}) Now move to the bottom right charge of the new polygon, $P_{n+1}$. Scatter this again via
\eq{
P_{n+1} \otimes D_n \rightarrow (7+3n,-2) \lor (3+n,-6-2n),
}{}
and choose $(3+n,-6-2n)$ for the same reason as before. This effectively grants us the enclosing light blue hexagon in \Fig{su3alg2}, so we are back to the shape we started with, but rescaled.  Crucially this hexagon contains a higher weight on the diagonal, which means $D_{n+1}$ is ready for use for step $n+1$.
\end{itemize} 

Similarly to the previous algorithm, the resulting infinite sequence of alternating hexagons covers $z=0$ fully.
From this construction, we automatically obtain arbitrarily large representatives of $z=1$ and $z=2$ by the process
\eq{
(n,n) \otimes (1,0) \rightarrow (n+1,n) \lor (n-1,n)
}{adj2su3}
Using charge conjugation, we can then obtain both channels. The action of the roots then moves charges towards the origin, allowing us to claim completeness of the full $SU(3)$ lattice.

\section{Representation Completeness}\label{app:reprcompl}

In the main text, we have demonstrated how a nonabelian symmetry group $G$ can be exploited to derive completeness in the abelian charge lattice of the Cartan subgroup $H$.  
These results suggest a natural follow-up question: do our assumptions also imply the existence of all possible irreducible nonabelian representations of the symmetry group $G$?   This line of inquiry is particularly well motivated in light of existing statements concerning completeness in the spectrum of irreducible representations~\cite{Polchinski:2003bq, Harlow:2018tng}. 

Unfortunately, completeness in the Cartan charge lattice does not, in and of itself, carry any direct implications about the spectrum of nonabelian representations.  The reason for this is straightforward.   While charge completeness certainly guarantees the existence of arbitrarily large nonabelian representations, this fact tells us absolutely nothing about the precise properties or structure of these representations. So any given abelian charge has the freedom to reside within any of an infinite number of arbitrarily large representations.

In the example of $SU(2)$ angular momentum, this observation corresponds to the fact that a state of a given $J_z$ can appear in any representation of spin  $J \geq |J_z|$. Without additional information, there is no way to ascertain the precise value of $J$.

In this section we briefly summarize some of our attempts at proving representation completeness. In Apps.~\ref{app:aaron}, \ref{app:so3}, \ref{app:so4}, our analysis will follow that of Ref.~\cite{Hillman:2024ouy}, which essentially used the full color structure of the dispersion relation in \Eq{DR} to deduce powerful constraints on the spectrum of nonabelian representations. There, our results will largely be negative.

We end with \Sec{app:channels}, in which we describe how, with an extra assumption, our algorithm succeeds in yielding representation completeness.

\subsection{Completeness from Proof by Contradiction}\label{app:aaron}

\vspace{-2mm}

The basic strategy of Ref.~\cite{Hillman:2024ouy} is to assume certain spectra and then act on \Eq{DR} with wisely chosen color projectors  to generate a contradiction \footnote{For earlier work bootstrapping ansatze for theories with color and gravity, see Ref.~\cite{Bachu:2022gof}.}.
 Let us now briefly review this approach.

Consider the scattering of a pair of particles in the nonabelian representations $R_1$ and $R_2$ of the symmetry group  $G=SO(N)$.  In terms of the four-point scattering amplitude, the external legs reside in the representations $R_1=R_4$ and $R_2=R_3$. The particles in the $s$ or $u$ channels will transform under representations appearing in the direct sum decomposition of $R_1 \otimes R_2$. We  denote the set of representations in this direct sum by $\mathcal{S}(R_1,R_2)$.

Given external states with the general nonabelian indices $i_1, i_2,i_3,i_4$, it is convenient to define the projectors $(\Proj_R^{I})^{i_1 i_2 i_3 i_4}$, for $I=s,t,u$, as a basis of color space.   Concretely, $(\Proj_R^{I})^{i_1 i_2 i_3 i_4}$ is the tensor structure induced by a state in the representation $R$ exchanged in each channel.  By crossing symmetry, we know that
$(\Proj_R^{s})^{i_1 i_2 i_3 i_4} = (\Proj_R^{t})^{i_3 i_2 i_1 i_4} = (\Proj_R^{u})^{i_1 i_3 i_2 i_4}$.
The projectors exhibit a natural scalar product,
$\langle \Proj_R^I,\Proj_{R'}^{I'} \rangle = (\Proj_R^{I})^{i_1 i_2 i_3 i_4} (\Proj_{R'}^{I'})^{i_1 i_2 i_3 i_4}$,
where the projectors are normalized to ensure that $\langle \Proj_R^I,\Proj_{R'}^{I} \rangle = \delta_{R R'}$. 

The dispersion relation in Ref.~\cite{Hillman:2024ouy} is similar to \Eq{eq:c2su}, but with $s$-dependent subtraction and expressed in terms of color projectors,
\eq{
\hspace{-2mm} - \frac{8 \pi G_N}{t}\!\Proj_{0}^{t} +\cdots {=}\! \sum_{R \in \mathcal{R} } \left[K_R(s,t)\Proj_R^s \,{+}\, K_R(u,t)\Proj^u_R \right]  ,\hspace{-2mm}
}{eq:proj1}
where $\mathcal{R} \subseteq \mathcal{S}(R_1,R_2)$ is the set of representations that are exchanged in the $s$ and $u$ channels. 
To make contact with Ref.~\cite{Hillman:2024ouy}, we define 
\hspace{-3mm}\eq{K_R(X,t) {=} \sum_J \int_{M^2}^{\infty} ds^{\prime} \frac{\mathbb{G}_J^{(D)}\!\!\left(1+\frac{2t}{s^{\prime}}\right) \text{Im}f_{J,R}(s^{\prime})}{\pi s^{\prime} (s^{\prime}+t)(s^{\prime}-X)},
}{}
with $\mathbb{G}_J^{(D)}$ the Gegenbauer polynomials. Here, for brevity we write the dispersion relation for massless external states, but introducing masses is straightforward. 

To derive a contradiction, we first assume some choice for $\mathcal{R}$.  Second, we contract both sides of \Eq{eq:proj1} with a tensor $v^{i_1 i_2 i_3 i_4}$.  This object is completely arbitrary, since we have the freedom to choose any external color polarizations.  Normalizing this tensor so that $\langle \Proj_0^t, v \rangle = 1$, we obtain the following equation,
\eq{
\hspace{-1.5mm} - \frac{8 \pi G_N}{t} \,{=} \! \sum_{R \in \mathcal{R} }\left[K_R(s,t)\langle \Proj_R^s, v\rangle {+} K_R(u,t)\langle \Proj^u_R, v \rangle\right] ,\hspace{-1.5mm}
}{}
dropping the subleading terms in Eq.~\eqref{eq:proj1} at small $t$. Now, if there exists some $v$ for which
\eq{
\langle \Proj_R^s , v \rangle = \langle \Proj_R^u , v \rangle = 0,
}{constraints3}
for all $R \in \mathcal{R}$, then \Eq{eq:proj1} cannot be satisfied and $\mathcal{R}$ is not a consistent spectrum. In such a case, there must exist in the spectrum at least one representation in $\mathcal{S}(R_1,R_2)$ that is not in $ \mathcal{R}$ in order for the dispersion relation to be consistent~\footnote{Our results on charge completeness can be understood in this language.  By scattering the charges $\vec q$ and $\vec q^{\,\prime}$, we have used that there is a state of charge $\vec q +\vec q^{\,\prime}$ or $\vec q -\vec q^{\,\prime}$, corresponding to the $s$ or $u$ channel.  This implies that there exists some choice of $v$ for which \Eq{constraints3} is satisfied where the set of exchanged representations $\mathcal{R}$ contains neither $\vec{q}+\vec{q}^{\,\prime}$ nor $\vec{q}-\vec{q}^{\,\prime}$.}.

Instead of looking for a general tensor $v$ that solves \Eq{constraints3}, the authors of Ref.~\cite{Hillman:2024ouy}  effectively insert the identity operator in the space of projectors $|\Proj_{R'}^t\rangle \langle \Proj^t_{R'}|$ to obtain
\eq{
\langle \Proj_R^s ,\Proj_{R'}^t\rangle \langle \Proj^t_{R'}, v \rangle &= \langle \Proj_R^u ,\Proj_{R'}^t\rangle \langle \Proj^t_{R'}, v \rangle = 0 \\
(M_{st})_{RR'}v_{R'} &= (M_{ut})_{RR'}v_{R'} = 0,
}{constraints4}
where we introduced the vector $v_{R'} = \langle \Proj^t_{R'}, v \rangle$ and the matrix $(M_{st})_{RR'} = \langle \Proj_R^s , \Proj_{R^{\prime}}^t \rangle$, and similarly for $M_{ut}$. Note that $(M_{st})_{R R^{\prime}}=(M_{ut})_{R R^{\prime}} (-1)^{g_{R^{\prime}}}$, where $g_{R^{\prime}}=0$ for symmetric representations and $g_{R^{\prime}}=1$ for antisymmetric ones.
The authors of Ref.~\cite{Hillman:2024ouy} then solved the linear system of equations in \Eq{constraints4} to find inconsistent sets of exchanged representations $\mathcal{R}$. They studied the scattering of fundamentals for both $SO(N)$ and $SU(N)$ and the scattering of adjoints for $SO(N)$. 

This approach is challenging to implement for arbitrary initial states and general groups for two reasons.
First, computing $M_{st}$ and $M_{ut}$ for general groups and representations is a challenging task. Second, the cardinality of $\mathcal{S}(R_1,R_2)$ increases with the rank of the initial representations $R_1$, $R_2$. Scanning over all consistent subsets of $\mathcal{S}(R_1,R_2)$ requires solving of order $2^{|\mathcal{S}(R_1,R_2)|}$ systems of linear equations.

\vspace{-2mm}

\subsection{$SO(3)$ Symmetry}\label{app:so3}

\vspace{-3mm}

For the symmetry group $G=SO(3)$, the matrices $M_{st}$ and $M_{ut}$ are known in closed form.  
In particular, these objects are the Racah $W$-coefficients, which up to a phase are the Wigner $6$-$j$ symbols that re-express angular momenta in the $t$ channel in terms of angular momenta in the $s$ and $u$ channel.  Explicitly, $M_{st}$ is proportional to
$\sqrt{(2J_{12}{+}1)(2 J_{23}{+}1)}W(j_1 j_2 j_4 j_3; J_{12} J_{23})$
and similarly for $M_{tu}$.  In the terminology of angular momentum,
 $j_1,j_2,j_3,j_4$ are the external spins and $J_{12}, J_{23}$ are the exchanged spins.

Using this expression, we were able to show that any set of exchanged representations $\mathcal{R}$ comprising a finite set of spins between $0$ and $N$ presents no inconsistency, which we have verified explicitly for $N\leq 4$.  If any spin between $0$ and $N$ is missing, then the spectrum is inconsistent.  The case of $N=1$, which includes spin-zero and spin-one states, was considered in Ref.~\cite{Hillman:2024ouy} and shown to be consistent.  
Importantly, this approach indeed indicates that an infinite number of representations is not needed for scattering consistency, since $N$ can be finite.

\vspace{-2mm}

\subsection{$SO(4)$ Symmetry}\label{app:so4}

\vspace{-3mm}

For $G=SO(4)$, we can also compute $M_{st}$ and $M_{ut}$ analytically.  This is possible because $SO(4) = SU(2)^2/\mathbbm{Z}_2$, which implies that we can derive the Racah matrices in $SO(4)$ by computing tensor products of Racah matrices of $SU(2)$ (which coincide with those of $SO(3)$). At the same time, given that completeness of $SU(2)$ cannot be proven using this method, it also cannot be proven for $SO(4)$, as discussed in \Sec{sec:SO4}.  

Something more can be said for the group $O(4)$. Each representation of $O(4)$ is of the form $(j_a, j_b)\oplus (j_b, j_a)$, where $j_a$ and $j_b$ are the spins relative to the two $SU(2)$ factors. It is simple to compute Racah matrices for scattering of representations such that either $j_a = j_b$ or $j_b = 0$. However, other than confirming what we already knew from the weight lattice argument, we did not obtain any new general results. In particular, our experiments suggest that representation completeness cannot be proven using this strategy. We checked that for all the scattering processes we could consider, we could always exclude from a set one arbitrary representation, suggesting that a set containing all the representations but one is consistent. This conclusion could have been anticipated by simply counting the number of degrees of freedom in \Eq{constraints4} and the number of equations.
In any case, a systematic study would be required to conclude that this approach cannot be used to prove completeness.

\vspace{-2mm}

\subsection{Completeness with Both Channels}\label{app:channels}

\vspace{-2mm}

If, for whatever reason, we are granted knowledge that {\it both} channels in \Eq{eq:c2su} are necessarily nonzero, then we can straightforwardly derive much stronger claims of completeness.   In this case, the scattering of charges $\vec{q}$ and $\vec{q}^{\,\prime}$ will necessarily entail new states of charge  $\vec{q} + \vec{q}^{\,\prime}$ and also  $\vec{q}-\vec{q}^{\,\prime}$.  Mechanically, we then deduce the existence of any charge from the sum or difference of other charges.
For $G=U(1)$, charge completeness follows trivially from the existence of even a single particle with $q=1$. Scattering $q=1$ with $q^{\prime}=1$ requires a new particle with charge $q+q'=2$, and so on.

This approach generalizes easily to $G=SO(3)$, assuming the spectrum contains a particle in the fundamental spin-one representation. Scattering a pair of such particles with $q=1$ and $q^{\prime}=1$, we obtain a particle with charge $q+q'=2$.  This state necessarily resides in the spin-two representation, since this is the only representation that carries this charge in the tensor product of spin-one with itself.  We then scatter states in the spin-one and spin-two representations with charges $q=1$ and $q'=2$, yielding a new particle with $q+q'=3$.  Iterating this process generates all possible spin representations.

The generalization to $SO(N)$ follows easily from \Eq{constraints3}. The requirement that both channels must be nonzero corresponds to enforcing only one of the two equations. By choosing $v\propto \Proj_{R'}^s$ the equation $\langle \Proj_R^s, v \rangle = 0$ is satisfied for all $R\neq R'$, meaning that $R'$ must be included in the spectrum. Given that $R'$ is any representation, representation completeness is proven. Importantly, we also have to check that $\langle \Proj_0^t, v \rangle$ is different from zero. This follows immediately from the completeness relation $\Proj_0^t \propto \sum_R c_R\Proj_R^s$, with $c_R\neq 0$~\cite{Birdtracks}.
 

\end{document}